\documentclass{aa}

\usepackage[varg]{txfonts}
\usepackage{graphicx}
\usepackage{url}
\usepackage[pdftex, colorlinks, linkcolor = blue]{hyperref}
\hypersetup{citecolor = blue}

\hypersetup{draft}
\begin{document} 

\title{The role of host star variability in the detectability of planetary phase curves}

\author{Hidalgo, D.\inst{1}$^,$ \inst{2}
		Alonso, R.\inst{1}$^,$ \inst{2}
		Pallé, E. \inst{1}$^,$ \inst{2}
        } 

\offprints{D. Hidalgo, \email{dhidalgo@iac.es}}

\institute{Instituto de Astrof\'\i sica de Canarias, 38205 La Laguna, Tenerife, Spain \and
			   Dpto. Astrof\'\i sica, Universidad de La Laguna, 38206 La Laguna, Tenerife, Spain
           }

\date{}


\abstract{Phase curves, or the change in observed illumination of the planet as it  orbits around its host star, help us to characterize their atmospheres. However, the variability of the host star can make their detection challenging: the presence of starspots, faculae, flares and rotational effects introduce brightness variations that can hide other flux variations related to the presence of an exoplanet: ellipsoidal variation, Doppler boosting and a combination of reflected light and thermal emission from the planet. Here we present a study to quantify the effect of stellar variability on the detectability of phase curves in the optical. On a first stage, we simulate ideal data, with different white noise levels, and with cadences and total duration matching a quarter of the \textit{Kepler} mission. We perform injection and recovery tests to evaluate the minimum number of planetary orbits that need to be observed in order to determine the amplitude of the phase curve with an accuracy of 15\%. We also evaluate the effect of a simplistic stellar variability signal with low amplitude, to provide strong constraints on the minimum number of orbits needed under these ideal conditions. On a second stage, we apply these methods to data from the quarter Q9 of the \textit{Kepler} mission, known for its low instrumental noises. The injection and recovery tests are performed on a selected sample of the less noisy stars in different effective temperature ranges. Even for the shortest explored planet period of 1 day, we obtain that observing a single orbit of the planet fails to detect accurately more than 90\% of the inserted amplitude. The best recovery rates, close to 48\%, are obtained after 10 orbits of a 1d period planet with the largest explored amplitude of 150 ppm. The temperature range of the host stars providing better recovery ratios is $5500 \text{K}<T_{eff}<6000 \text{K}$. Our results provide guidelines to select the best targets in which phase curves can be measured to the greatest accuracy, given the variability and effective temperature of its host star, which is of interest for the upcoming \textit{TESS}, \textit{CHEOPS} and \textit{PLATO} space missions.
}


\maketitle


\section{Introduction}
Studying planetary systems which transit their host star provides a great opportunity to derive their physical properties. During a \textit{transit} one can measure the exoplanet radius $(R_p)$ in units of the stellar radius $(R_*)$ through the depth of the transit $\sim (R_p/R_*)^2$, the scaled semi-major axis, the impact parameter \citep{Mandel2002} and the orbital period. If the geometry (inclination and eccentricity) of the system is adequate, a \textit{secondary eclipse} can also be observed. Along one orbital period, the time-dependent change in the brightness of the light curve measured due to the reprocessed light of the exoplanet from its host star is known as the \textit{phase curve}. This change in brightness is determined by the combination of emitted and reflected light from the host star in a particular bandpass \citep{Cowan2011}.

Secondary eclipses and phase curves can provide a wealth of information on a given planet and help characterize its atmosphere. The depth of the secondary eclipses gives us clues to constrain the albedo of the planet \citep{Angerhausen2015}, while the timing and duration help us to determine the orbital parameters \citep{Agol2017}. A detailed study of phase curve can tell us, for instance, how efficient is the energy redistribution between the exoplanet's dayside and nightside, or how atmospheric bright spots can shift due to persistent wind patterns (e.g. \cite{Knutson2007}, \cite{Demory2013}, \cite{Armstrong2016}). All these effects can be characterized in different wavelengths in order to show the vertical temperature structure and also the chemical composition of the exoplanet atmosphere.

Phase curves have been observed in the near- and mid-infrared using the \textit{Spitzer Space Telescope} for several planets, such as HD 209458b \citep{Zellem2014}, the hot Saturn HD 149026b \citep{Knutson2009} or WASP-43b \citep{Stevenson2017}. With the launch of \textit{Kepler Space Telescope} \citep{Borucki2008}, phase curves have also been observed in the optical (e.g. Kepler-7b, \cite{Esteves2013}). Only one phase curve in the optical band had been previously observed: CoRoT-1b \citep{Snellen2009}. In the optical band, the planet-star contrast is much lower, and the contributions from ellipsoidal variations or the beaming effect (also named Doppler boosting) become important (e.g. \cite{Loeb2003}, \cite{zucker2007}, \cite{Shporer20017}). It should also be noted that in this bandpass, the phase curve is degenerated. In the extreme cases of hot Jupiters there is a component of the planet's own emission that contaminates the signal, leaking into the \textit{Kepler} bandpass (e.g. \cite{Morales2007}, \cite{Heng2013}). For further insight into observations and theory of phase curves see \cite{Parmentier2017}.

High-precision space photometry is a powerful tool for the study of transits and secondary eclipses as well as phase curves. However, aside from the systematic variations of photometry due to telescope movement and other electronic processing, the intrinsic variability of the star itself plays a fundamental role when performing data reduction and analysis.

 The Sun has been studied in depth to unravel the mechanisms of stellar activity emerging from the surface \citep{Berdyugina2005} and this activity can also be seen in other stars. Dark starspots corresponding to concentrations of the magnetic field emerging from the photosphere, while bright faculae correspond to the enhanced network magnetic field, which is dispersed over a much larger area \citep{Kitiashvili2013}. In general, stellar variability usually shows periodicity. However, in many cases, this periodicity may vary due to the fact that the configuration of the magnetic field on the star's surface changes rapidly over time and results in irregular variability in the light curve \citep{Brown1994}.

\begin{figure}
  \resizebox{\hsize}{!}{\includegraphics{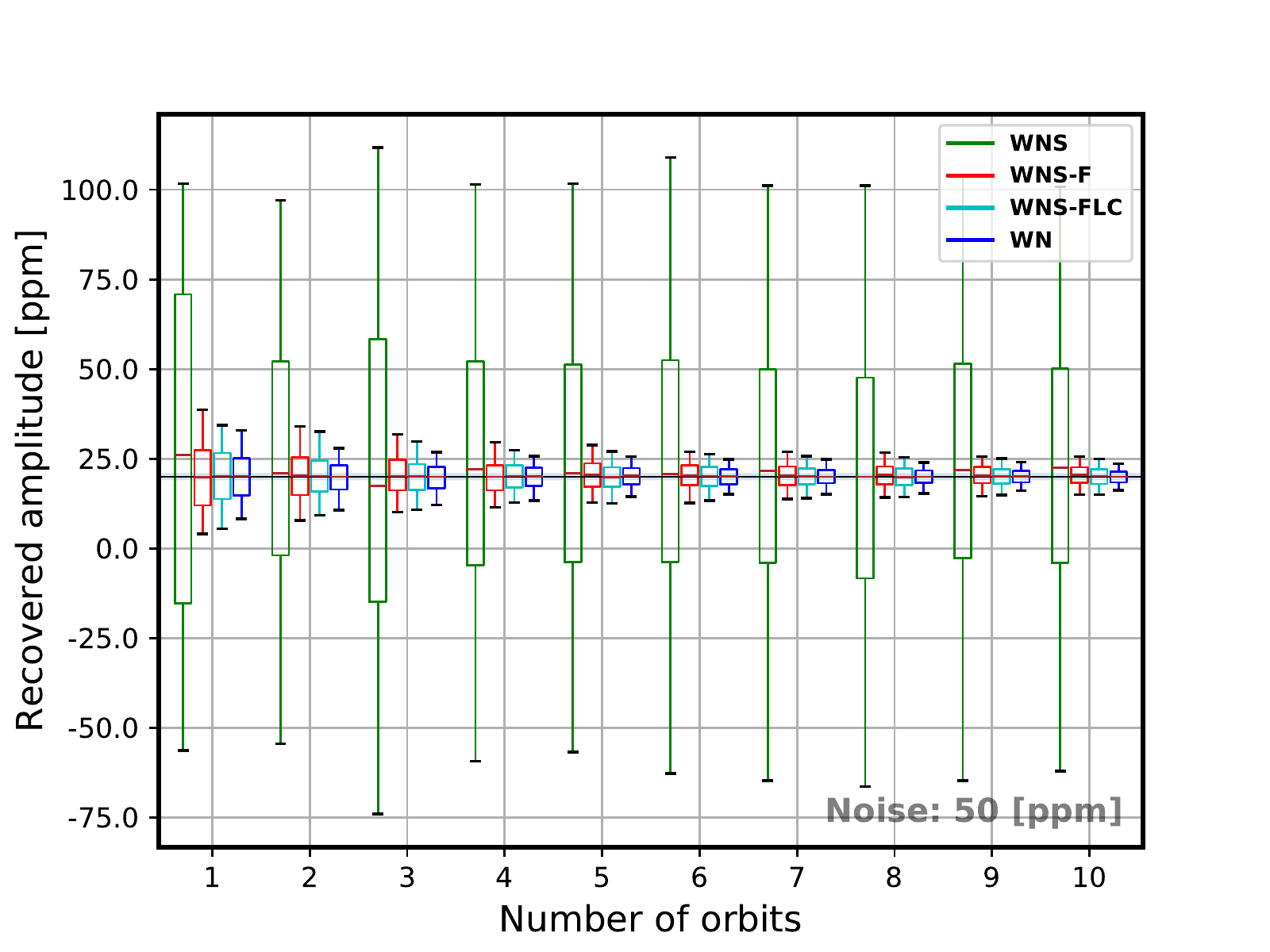}}
  \resizebox{\hsize}{!}{\includegraphics{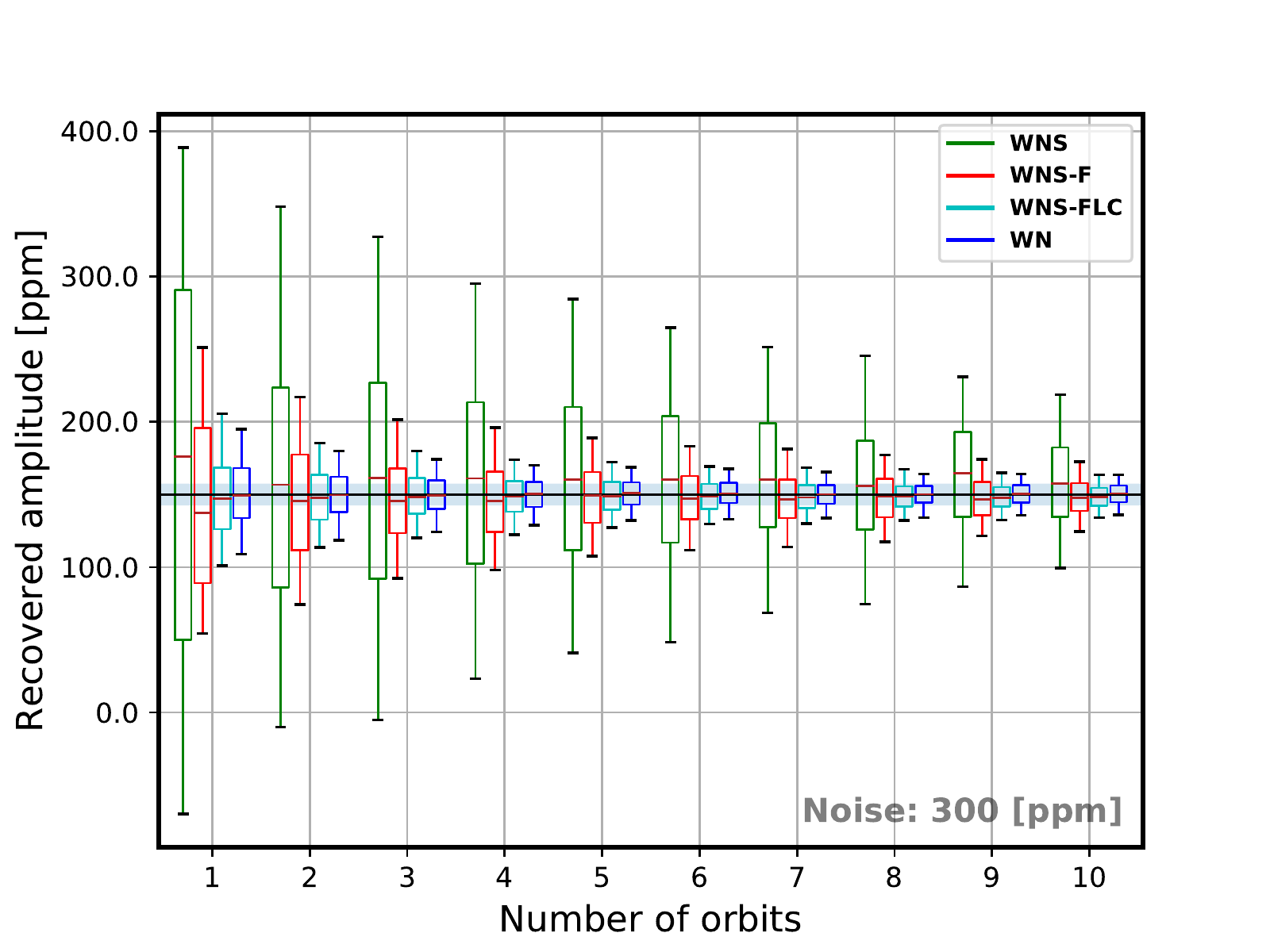}}
  \caption{Results of the recovery of the amplitude of phase curves versus the number of orbits with which the fit has been made. Each plot represents an example light curve generated with different white noise and amplitude (top: noise of 50 ppm per exposure, amplitude of 20 ppm, and period of 2.5 days, and bottom: noise 300 ppm per exposure, amplitude of 150 ppm, and period of 7.5 days) for an exposure time of 29.4 minutes. The horizontal black line indicates the amplitude of the entered phase curve and the shaded area in blue is the range in which the phase curve detection criterion is valid. For each combination number of orbits, there are 4 boxplots that indicate the 4 different cases considered: \textit{WN} (blue) \textit{WNS} (green), \textit{WNS-F} (red) and \textit{WNS-FLC} (light blue-green). The red horizontal line on each boxplot indicates the median, the rectangular box is the IQR (see definition on section \ref{subsec:metVT}) and the thin error bars span the whole distribution of points. In the top figure there is no detection, as the IQR is not fully inside the blue-shaded area, whereas in the bottom figure there is a detection when we accumulate 7 or more orbits}.
  \label{fig:WN_re}
\end{figure}

\cite{Luisa2017} have published a study focusing on the distinction of exoplanet albedo from stellar activity. Unlike their study which focuses on the fit of stellar variability to obtain an albedo estimation, requiring a continuous time interval of data for at least one stellar rotation, the study we present here focuses on fitting the phase curve, without the need for these phase curves to be consecutive. This is the case when the scheduling of long observations is not easily possible (e.g. repetition and combination of \textit{HST} phase curves, \textit{K2} re-observations of the same star) or observations that are part of a tight schedule of different programs (e.g. \textit{CHEOPS} or \textit{JWST}).

For the new generation of space-based telescopes such as \textit{TESS} \citep{Ricker2015}, \textit{CHEOPS} \citep{Broeg2013} and \textit{PLATO} \citep{Rauer2014}, it is important to determine the phase curve detection limitations due to stellar variability in order to prepare the exploitation of their data.

In this paper, we will determine the detectability of optical phase curves of exoplanets using the \textit{Kepler} database \citep{Jenkins2010}. In section \ref{sec:VT}, we perform simulated phase curve injection tests and describe the analysis procedure. In section \ref{sec:Kep}, we apply the same methodology to \textit{Kepler} data. We distinguish between two cases: one in which we attempt to filter out the stellar variability and another case in which we do not, since in many cases we do not have enough information to perform the filtering. Finally, in section \ref{sec:sum}, we discuss the results and present our conclusions.


\begin{figure}[!h]
  \resizebox{\hsize}{!}{\includegraphics{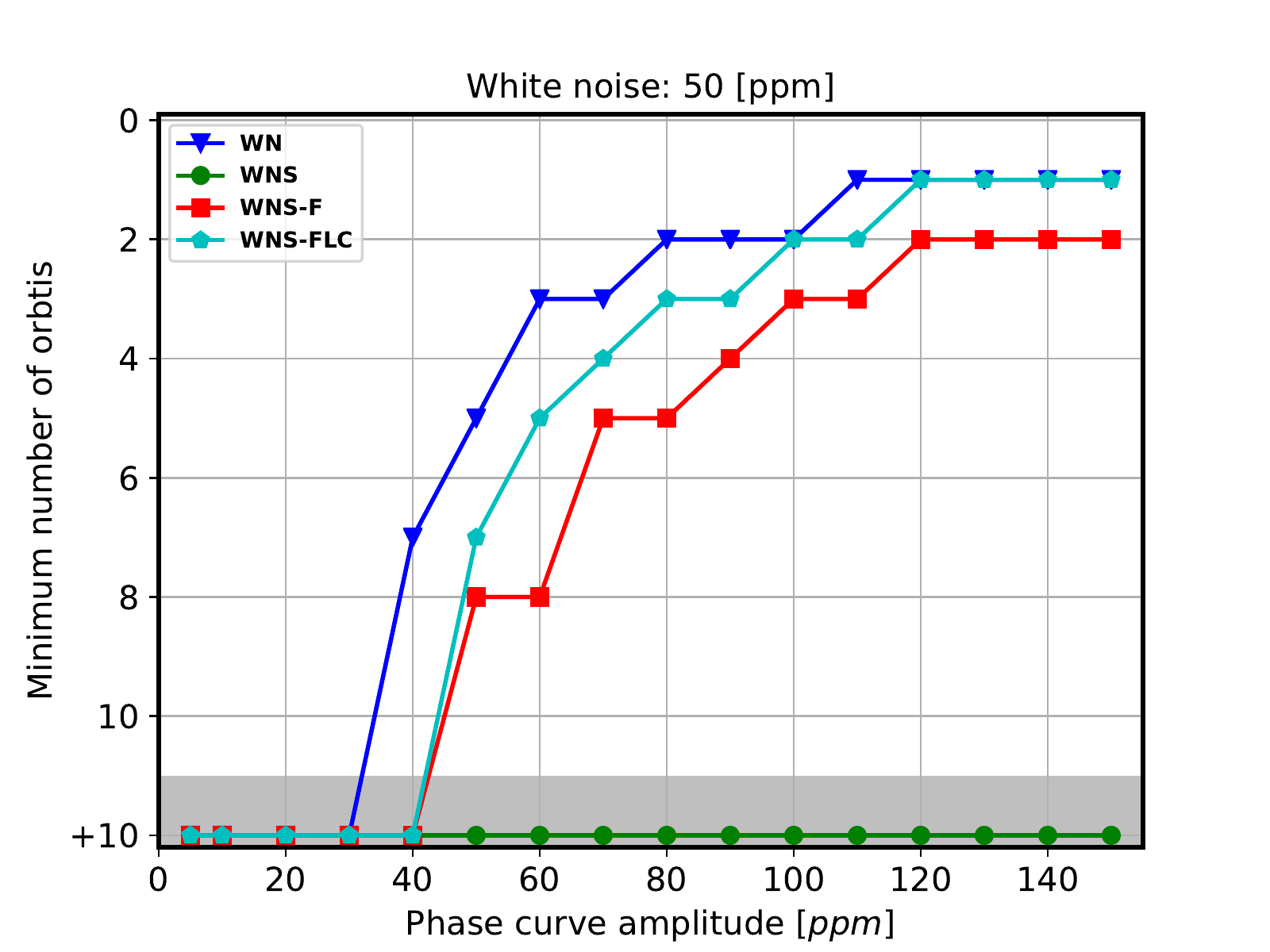}}
  \resizebox{\hsize}{!}{\includegraphics{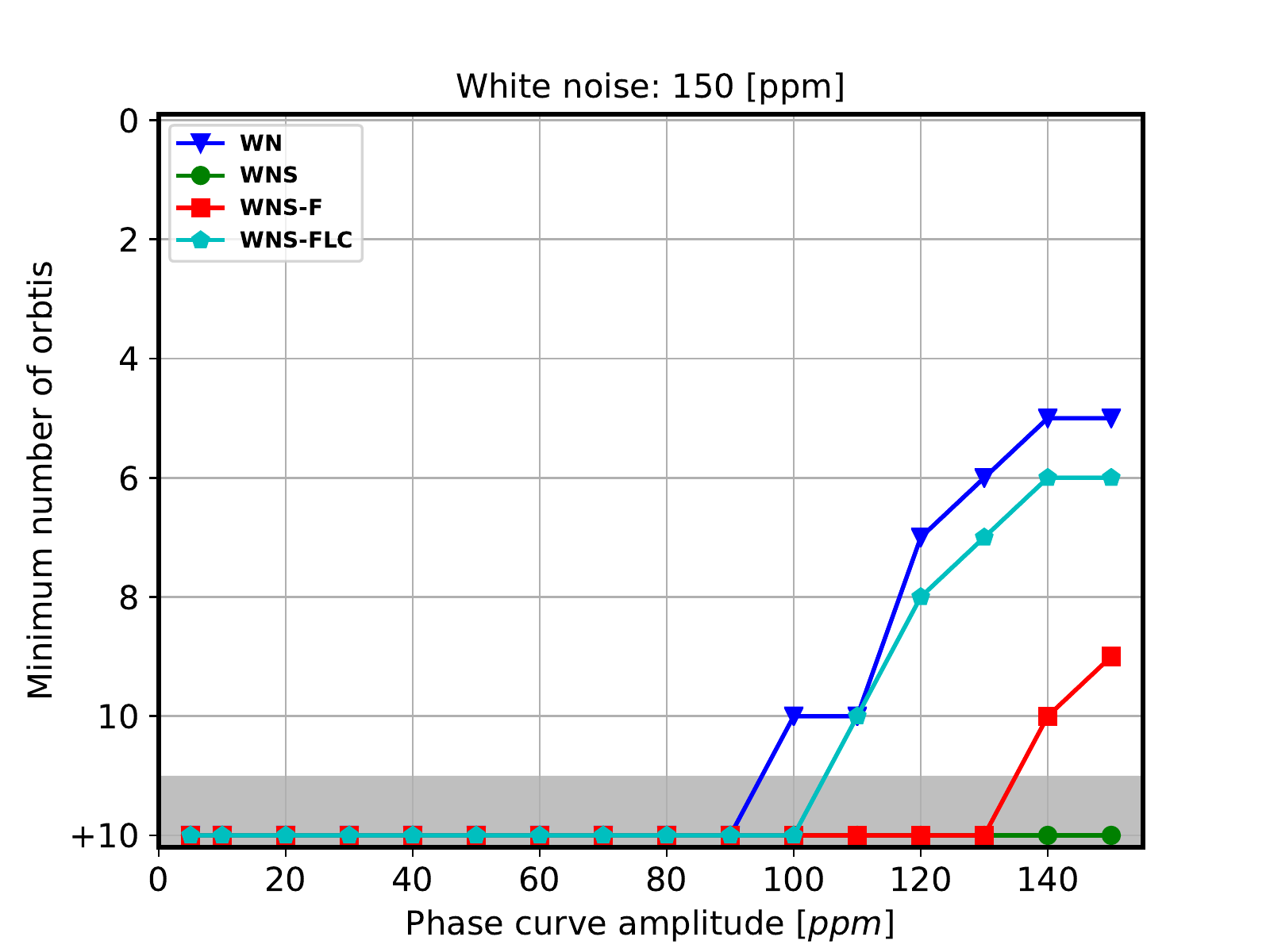}}
  \caption{Minimum number of orbits required to have significant phase curve detection. Each of the plots represents an initial white noise level (top: 50 $ppm$, bottom: 150 $ppm$) for a period of 2.5 days. For each of the different white noise curves 4 different cases are studied, as explained in Figure \ref{fig:WN_re}. The grey shaded area represents the case where more than ten orbits (+10) are required to perform a phase curve detection. The plot with an amplitude of 300 $ppm$ is not presented here since it is necessary more than 10 orbits in all cases to detect the phase curve.}
  \label{fig:WN_orb}
\end{figure}

\section{Phase-curve injection tests on simulated data}\label{sec:VT}
Through this manuscript, we define one orbit as the entire temporal interval that goes from the exit of a primary transit to the beginning of the next transit. As the goal is to study the influence of stellar variability that can not be corrected, we will adopt several ideal assumptions that explore the \emph{minimum} effect of this variability: 
\begin{itemize}
\item[$\bullet$]{The phase curve is injected and modeled by a simple sinusoidal. Higher harmonics such as tidal modulations are not included in this study.}
\item[$\bullet$]{The period and orbital phase are exactly known, and we will only study the effects on the amplitude of the sinusoid.}
\end{itemize}
Any departures from the previous assumptions will increase the detrimental effects of the stellar activity on the phase curve determinations.
In our study, the amplitude of the phase curve is defined as the peak to peak difference. Our injected light curves are sinusoidal signals shifted by $\pi / 2$ (so that the maximum is in phase 0.5). The amplitude of the phase curve is thus twice the sinusoidal semi-amplitude.

\subsection{White noise injection}
Since we will systematically introduce simulated phase curves in the form of sinusoidal signals into \textit{Kepler} light curves in the next section, we first use simulated light curves to check that the method gives consistent results. We decided to use \textit{Kepler}'s long cadence (LC) time steps of 29.4 min to unify criteria with the data used in section \ref{sec:Kep}. For our mock light curves, we also choose a duration equal to the quarter 9 (Q9) of the original mission \textit{Kepler}. With only one quarter we can perform our study, and the quarter 9 has been chosen because it has the lowest noise levels \citep{Howell2016}.

We generate white noise light curves with a normal distribution centered on 1.0, as if it were normalized, and with a standard deviation of 50, 150 and 300 ppm per exposure and for an exposure time of 29.4 minutes. The width of the distribution has been chosen taking into account the expected photometric performances of the next generation of space missions: \textit{TESS}, \textit{CHEOPS} and \textit{PLATO}. For the \textit{CHEOPS} mission, we would obtain this precision, with magnitudes in the V-band of $\sim$9.6, slightly more than $\sim$12 and $\sim$13, respectively, according to the exposure time calculator\footnote{\href{https://www.cosmos.esa.int/web/cheops-guest-observers-programme/open-time-workshop-2017}{https://www.cosmos.esa.int/web/cheops-guest-observers-programme/open-time-workshop-2017}} from \textit{CHEOPS} guest observers program. For \textit{TESS}, these magnitudes correspond to $\sim$10 and 8.5 for the case of 300 and 150 $ppm$, respectively\footnote{\href{https://heasarc.gsfc.nasa.gov/cgi-bin/tess/webtess/wtm.py}{https://heasarc.gsfc.nasa.gov/cgi-bin/tess/webtess/wtm.py}}. \textit{TESS} telescope won't be able to reach the estimated photometric error of 50 $ppm$, since the mission assumes 60 $ppm$ of background noise on hourly timescales \citep{Ricker2015}. \textit{PLATO} is still under development but a first estimation can be made \citep{Rauer2016} and it corresponds approximately to $\sim$11, $\sim$13.5 and $\sim$15, respectively.

\begin{table}
\caption{Examples of typical amplitudes of phase curves measured in \textit{Kepler} data, taken from \cite{Esteves2013}.}
\label{table:amplitude}
\centering
\begin{tabular}{cc}
\hline\hline
 & Amplitude $ppm$\\
\hline
Kepler 5b & $19.3^{+6.3}_{-5.3}$ \\
Kepler 7b & $48\pm 13$ \\
Kepler 12b & $22.9^{+4.3}_{-4.1}$ \\
Kepler 43b & $71\pm 44$ \\
Kepler 76b & $106.9^{+4.3}_{-4.4}$ \\
KOI-13b & $150.4\pm 2.7$ \\
HAT-P-7b & $73.3\pm 2.7$ \\
TrES-2b & $4.1^{+1.1}_{-1.0}$ \\
\hline
\end{tabular}
\end{table}

\subsection{Stellar variability and phase curve injection}
Our next step is to introduce stellar variability into our mock light curves. The stellar variability in \textit{Kepler} light curves has already been studied: as shown in the histogram in Figure 4 of \cite{Basri2011}. In both cases, for those stars that present periodicity in their variability and for those that do not, the most frequent amplitude of variability is around 2000 ppm. And their Figure 7 shows that the most frequent period for stars with variability around 2000 ppm is 14-15 days. If we take into consideration these results, a typical range for the amplitude of stellar variability would be between 100 and 15$\times 10^{3}$ ppm and periods between 8 and more than 14 days for that range, respectively. We assume a relatively favorable scenario for the detection of phase curves by introducing in our light curves stellar variability through a sinusoidal signal with an amplitude of 500 ppm and a period of 25.6 d (like the Sun). 

We simulate an idealized effect of stellar variability by introducing a second sinusoidal signal with a specific period and amplitude. We introduce periods distinctively shorter than the period of stellar variability, namely: 1.0, 2.5, 5.0 and 7.5 d. As for the amplitude of the phase curve, we explore the values: 5 ppm and from 10 to 150 ppm in steps of 10 ppm. These values span the typically detected amplitudes ranges for known exoplanets (see in Table \ref{table:amplitude}).


\begin{figure*}[h!]
\centering
   \includegraphics[width=17cm]{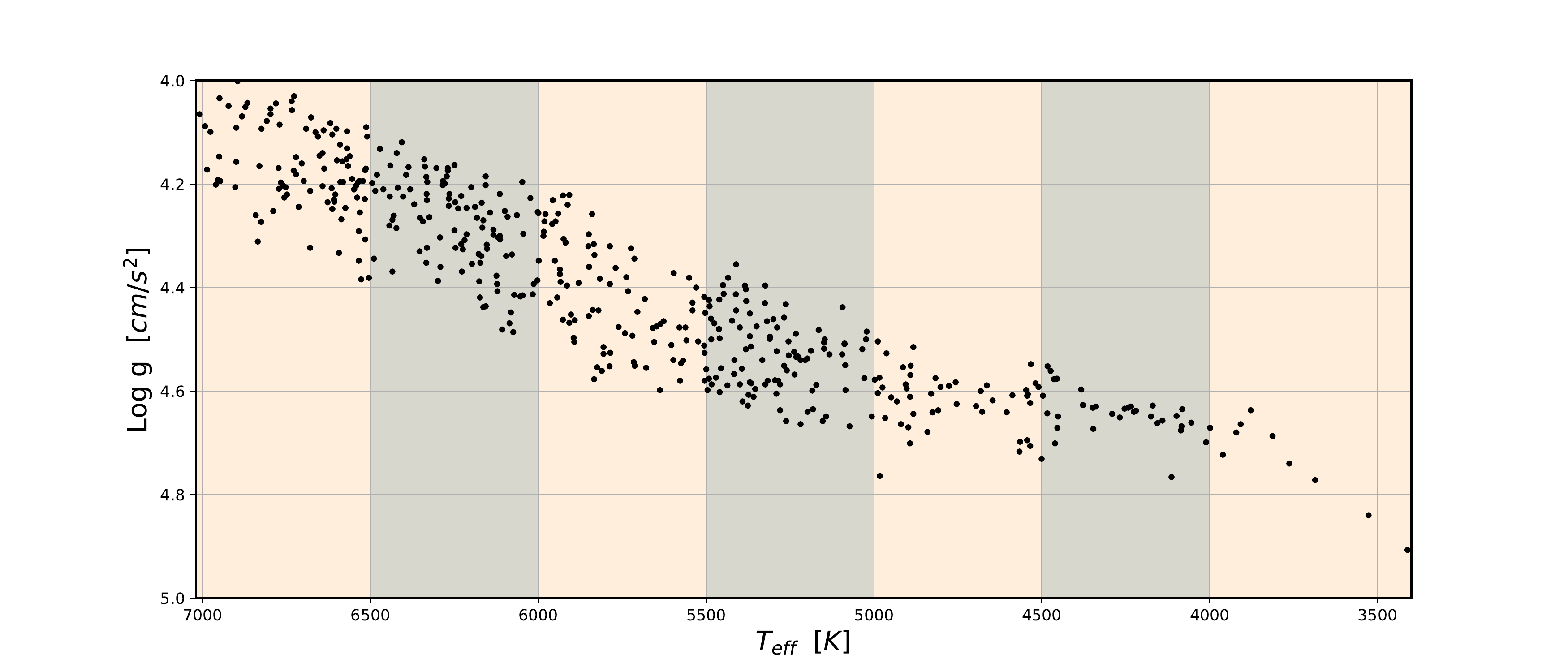}
     \caption{The diagram represents the logarithm of gravity versus effective temperature for the entire sample of stars used in this study, taken from \textit{Kepler Input Catalog} (\textit{KIC}, \cite{Brown2011}). The different shaded zones correspond to the effective temperature ranges in which the sample has been subdivided.}
     \label{fig:Kep_all}
\end{figure*}

\subsection{Light Curve analysis}\label{subsec:metVT}
Our mock light curves have a duration of 97 days, covering a significant amount of phase curves depending on the period we are using: 1.0, 2.5, 5.0 or 7.5 days. Phase curves cannot always be continuously observed from space due to mission planning constrains. For example, \textit{CHEOPS} will only observe phase curves individually. 

For this reason, we combine sub-sample datasets of our mock light curves, covering up to a maximum of 10 non-consecutive, randomly chosen phase curves. The combination is done in order to increase the signal to noise and to be able to better fit the simulated phase curve function. All fits are made using non-linear least-squares minimization. 

While recovering the phase curves, we explored the following cases:

\begin{enumerate}[$\bullet$]
\item \textit{WN:} Light curves containing only white noise + the injected phase curve.
\item \textit{WNS:} Same as in the previous case, with the added effect of stellar variability, but without filtering that variability.
\item \textit{WNS-F:} Same as before, but filtering stellar variability locally. This local filtering consists of taking each orbit individually and normalizing it by the median. In addition, since orbits can be found anywhere on the curve, we eliminate the local slope of an orbit by fitting a (straight) line.
\item \textit{WNS-FLC:} The light curve contains again stellar variability, and here we apply a detrending filter to the whole light curve to try to remove it. This filter consists of performing binning so each orbit is represented by a point and then, an interpolation fit function by third order splines.
\end{enumerate}

To retrieve the statistical significance of our study, we perform multiple combinations of different orbits in different parts of the light curve, so we can construct a distribution of the results.

The combination of 2 or more phase curves gives rise to a large number of possibilities. For each of the cases (from 2 to n orbits combinations, but we limit our study to 10 orbits), we make 100 random combinations of individual orbits. We also establish a criterion to determine if a phase curve has been properly detected: when the distribution of amplitudes obtained with all orbit combinations has both its quartile 3 and quartile 1 completely included in the amplitude range from -15\% to +15\% of the injected amplitude, and this injected amplitude is in the interquartile range (IQR) of the distribution.

\subsection{Results of phase curves recovery}\label{subsec:resWN}

We have applied the method explained in the previous section to different curves generated with 3 levels of white noise, with standard deviations of 50, 150 and 300 ppm. A box plot analysis of independent light curves is shown in Figure \ref{fig:WN_re}.

Following similar analysis for all light curves, the final recovery results can be seen in the Figure \ref{fig:WN_orb}, except for the 300 ppm case that did not produce any detection. The ideal case (blue line), it is the most favorable upper limit. On the other hand, we have the most realistic case, in which the light curve has stellar variability with no corrections (green line) the lower limit. In the middle (red and light blue line), we have the results for different methods of detrending the stellar variability.

The recovery of the injected phase curve is possible in a relatively low number of orbits for low values of white noise or large phase curse semi-amplitudes, for the idealized case of no stellar variability, or when this stellar variability is corrected over the full light curve duration. When one cannot make a correction of this variability, no detection of any phase curve that we have introduced can be accomplished with less than 11 orbits (many more will be needed for most cases). This will be the situation for \textit{CHEOPS} where the time interval of observation will not allow to deduce a trend in the variability of the star.

In the case of 50 ppm white noise assumption, we would be able to detect phase curves with amplitudes greater than 60 ppm for the WNS-FLC case or 70 ppm for the WNS-F, if we had at least 5 accumulated orbits with a period of 2.5 day. Conversely, for the 300 ppm white noise assumption, we cannot detect any phase curve, with period 2.5 days, having accumulated less 10 orbits (not shown). In the case of 150 ppm white noise, we are in an intermediate situation: we are able to detect amplitudes of the phase curve of more than 130 ppm in the WNS-FLC case or 140 ppm in the WNS-F.

\section{Application to Kepler data set}\label{sec:Kep}
In order to carry out this study with real photometric data, we need a minimum requirement: A time series long enough and a signal-to-noise small enough to be able to detect the amplitude of the phase curve. The \textit{Kepler} dataset is perfectly suited to these very specific requirements. Therefore, a number of stars from \textit{Kepler}'s complete star catalog, taken directly from \textit{NASA Exoplanet Archive}\footnote{\href{https://exoplanetarchive.ipac.caltech.edu/}{https://exoplanetarchive.ipac.caltech.edu/}}, have been selected to build the sample. The set of stars selected have been downloaded from the database from \textit{Mikulski Archive for Space Telescopes}\footnote{\href{https://archive.stsci.edu/kepler/data_search/search.php}{https://archive.stsci.edu/kepler/data\_search/search.php}} (MAST). For this study we used the light curves generated with the latest \textit{ Kepler Data Release 25}, more specifically, the flux vector used was created doing simple aperture photometry (SAP) with a precise artifact mitigation included called Pre-search Data Conditioning \textit{PDC\_SAP} \citep{Smith2012}. 

Among all the stars in Kepler's catalog, we reject the red giants and we consider only main sequence stars with an effective temperature lower than 7000 K (where most hot Jupiters have been found) and with a \textit{Kepler} magnitude brighter than 12 (to avoid instrumental noises on fainter targets). We divide the effective temperature range into different sub-ranges of 500 K and, in each sub-range. We select 100 stars, that have the smallest CDPP in 1.5 hours. Sub-ranges between 3500 K to 5000 K contain only 17, 37 and 48 stars, respectively.

We remove the giant stars by limiting the logarithm of gravity ($logg$). This limit varies from one temperature sub-range to another. Therefore, we take the greater $logg$ value ($max_{logg}$) for the each temperature and manually remove all stars that have an approximate value above $max_{logg}$ - 0.3. In Figure \ref{fig:Kep_all} we plot our sample of stars as a function of stellar effective temperature $T_{eff}$. Each shaded zone represents a sub-range in temperature.

To inject planetary signal in our sample of selected \textit{Kepler} light curves, we proceed as described in Section \ref{sec:VT}. Since these are real light curves, they already contain stellar variability signals and white noise. We extract the phase curve, using non-linear least-squares minimization and use the same IQR criteria for detection.

\subsection{Results}\label{subsec:resKep}
\begin{figure}[!]
	\resizebox{\hsize}{!}{\includegraphics{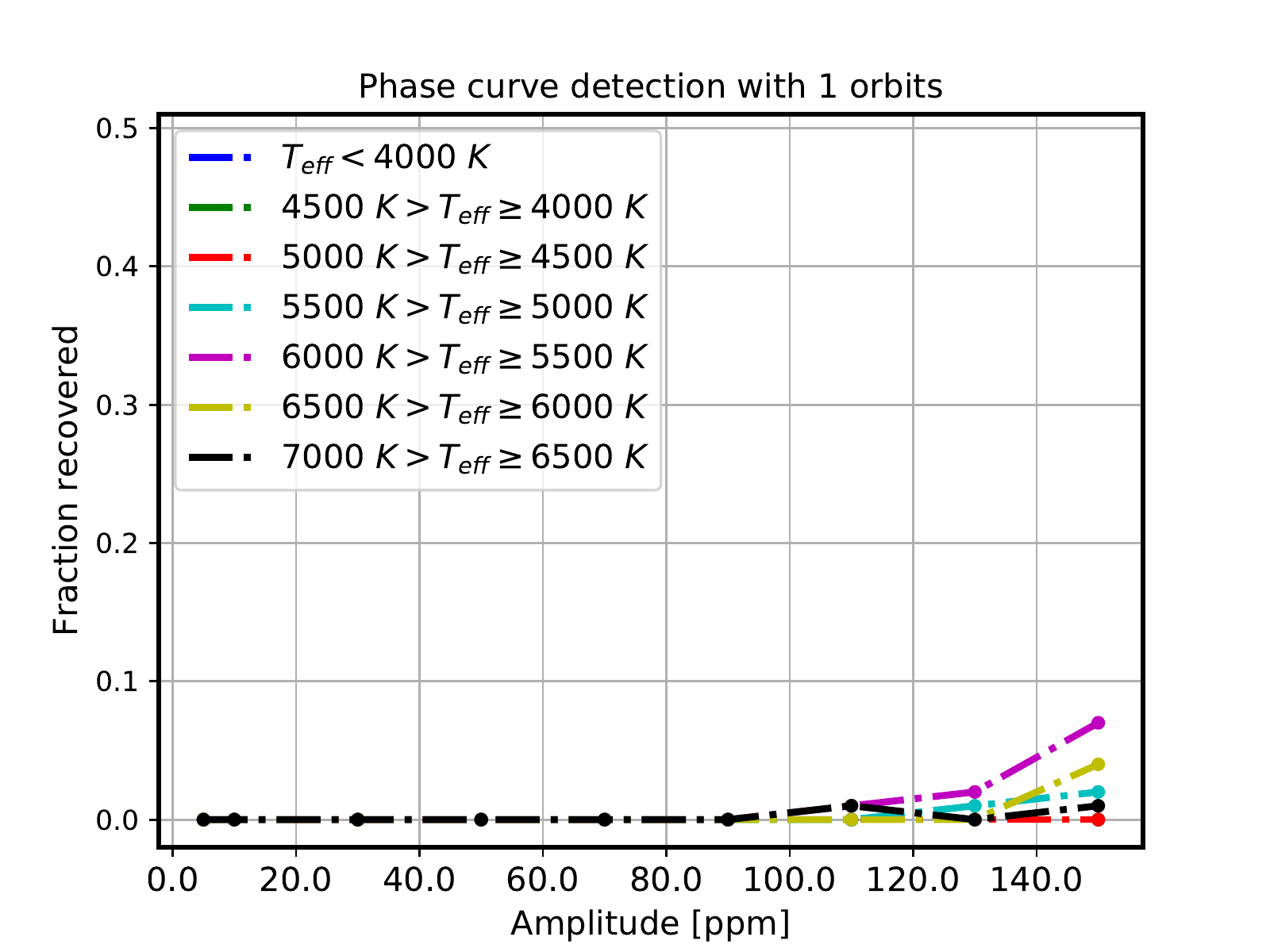}}
	\resizebox{\hsize}{!}{\includegraphics{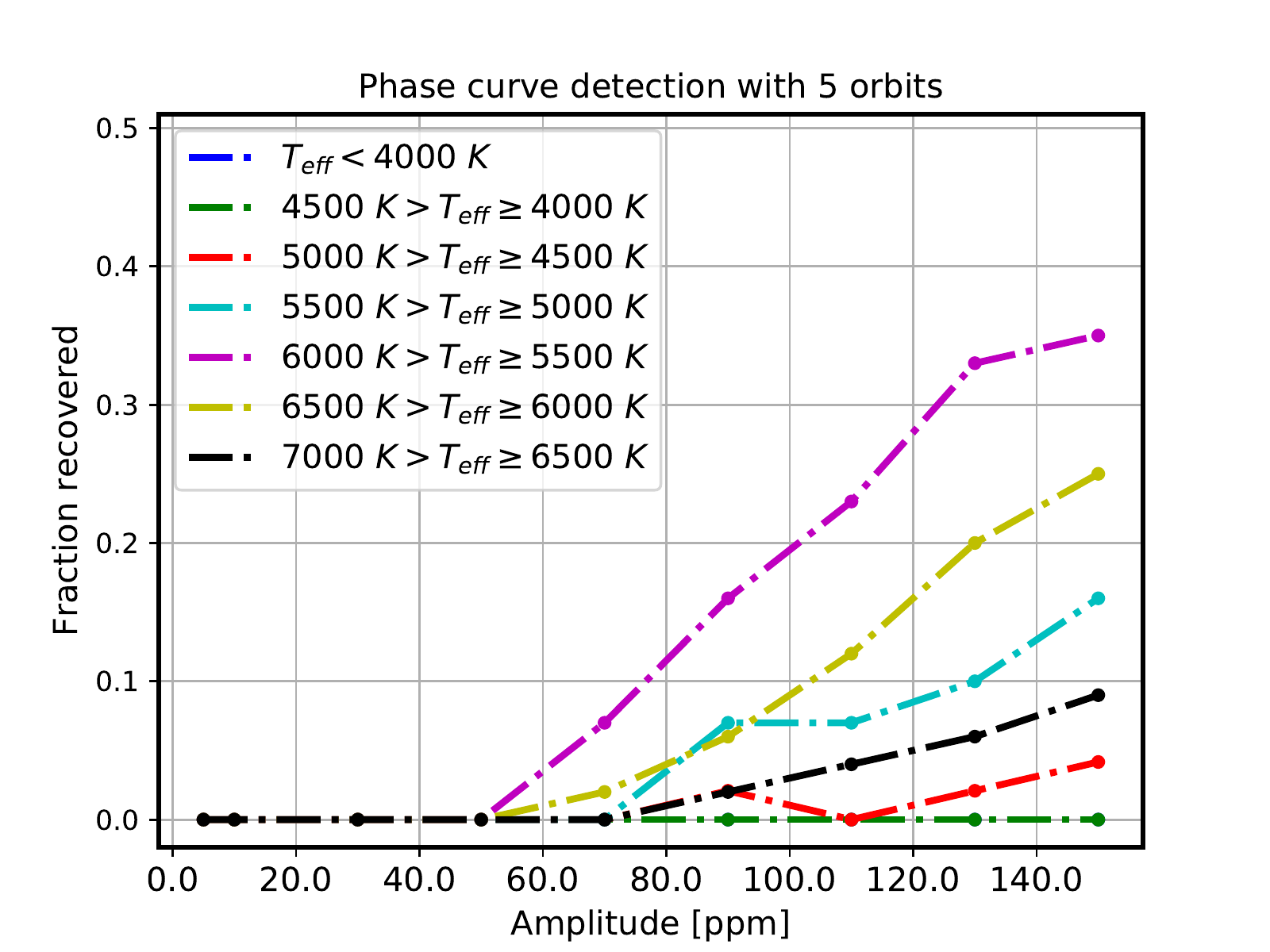}}
	\resizebox{\hsize}{!}{\includegraphics{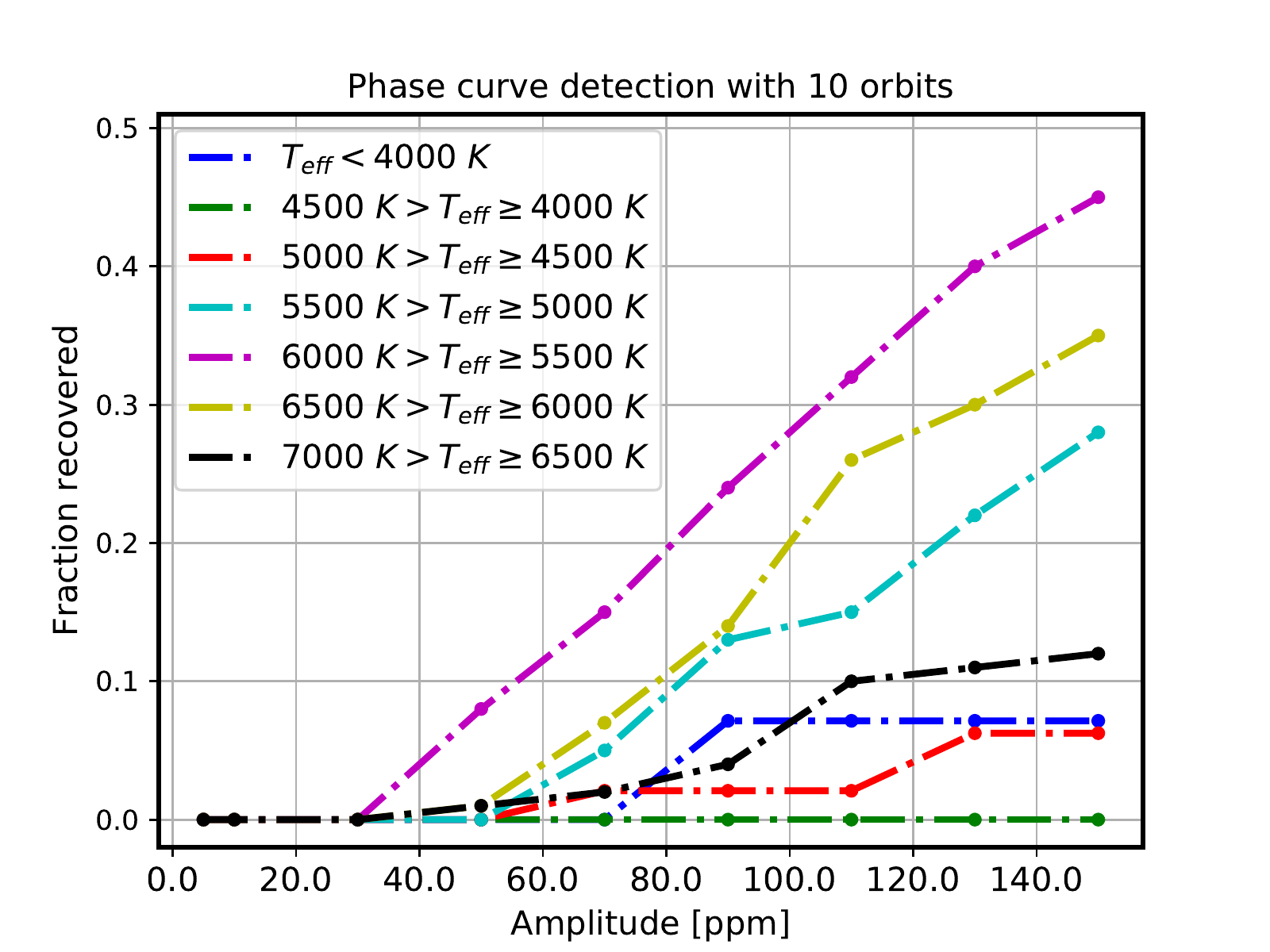}}
    \caption{Fraction of phase curves recovered as a function of the amplitude of the phase curve entered for a period of 2.5 days due to 1, 5 and 10 orbits considered (plots top, middle and bottom, respectively). Each of the colors of the individual dotted lines represents a different sub-range of effective temperature.}
  	\label{fig:Kep_N}
\end{figure}

Figure \ref{fig:Kep_N} shows the fraction of phase curves with the correct amplitude recovered for an input period of 2.5 days. Equivalent plots for other input periods (1, 5 and 7.5) are shown in Appendix \ref{app:0}. In each figure, we show the combination of 1, 5 and 10 orbits, (top, middle and bottom panels, respectively). As expected, the larger the amplitude of the injected phase curve, the greater is the probability that the phase curve is recovered. The probability also increases as one accumulates more orbits. For a period of 2.5 days, Figure \ref{fig:Kep_N} shows how the fraction of recovered phase curves is better as we accumulate orbits. In all cases, a phase curve is more easily detected when the star is of spectral type $G$ (solar type).

With one orbit, the fraction of phase curves is barely recovery for amplitudes even larger than 130 $ppm$. Independently of the period and number of orbits, the best result is for sun-like stars (spectral types $G$, from 5500 to 6000 K), but for one orbit we have a probability of recovering the phase curve below 8\%. For the rest of spectral types the fraction of recovered phase curves is below 2\% or null.

When 5 orbits are accumulated, the fraction of phase curves significantly increases. For $G$ type stars, the probability of recovered is below 35\%. For $G - F$ spectral types (yellow dotted line) this probability is around 20\%. And for the rest of spectral types, it is below the 10\% recovery factor.

If we accumulate up to 10 orbits, the detection rate increases but not significantly. For $G$ spectral type stars this probability reaches 45\%. For $F - G$ type stars the probability now is around 30\% and for the rest of spectral types barely cross the line of 10\%.

We perform the same type of study for the rest of the proposed periods (1.0, 5.0 and 7.5 d), obtaining similar results. In all cases, the behavior is similar. The $G$ spectral type is the one with the best fraction of phase curve recovery. Independently of the inserted period, using only observations during one orbit is not enough to correctly obtain the amplitude of the phase curve on 8\% of cases, in the most favorable scenario. When $G$ stars have 5 orbits, a fraction of $\gtrsim 40$\%, $\sim 28$\%, $\gtrsim 10$\% is obtained for a period of 1.0, 5.0 and 7.5 d, respectively. With 10 orbits, the behavior is similar but with higher fraction, $\gtrsim 50$\%, $\sim 38$\%, $\gtrsim 30$\% for a period 1.0, 5.0 and 7.5, respectively. For the rest of spectral types, we find a behavior similar to that of Figure \ref{fig:Kep_N}. The fraction of stars below 5500 K in which the phase curve amplitude was correctly detected is much smaller, in most cases not reaching the barrier of 10\%. For stars warmer than 6000 K, the pattern is also similar. Having 10 accumulated orbits, there is a larger fraction of phase curves recovery, around 15\% for periods below 5.0 d and 10\% if the period is 7.5 d.

We note that the longer the period, the sooner the phase curves begin to be observed. For a period of 7.5 d and 10 accumulated orbits, phase curves is recovered with an amplitude of 30 ppm. In the case of a 1.0 d period, that amplitude increases up to 50 ppm.

\section{Summary and conclusions}\label{sec:sum}
In this work, we have studied the role of host star variability in the detectability of planetary phase curves splitting it up into two parts. First, we have applied a method based on sinusoidal signals to simulate both the stellar variability and the phase curve of an exoplanet. We have assumed 4 different cases: \textit{WN}, a light curve only composed of white noise and the phase curve itself. \textit{WNS}, the same as above but adding a star variability of 500 $ppm$, \textit{WNS-F}, when we try to eliminate such variability locally and \textit{WNS-FLC} if we remove variability globally. Our results exposed in sub-section \ref{subsec:resWN} show that only in the most favorable cases of white noise plus a sufficiently large phase curve amplitude, we can consider that the detection of the phase curve of a planet is positive, taking into consideration the criterion defined in sub-section \ref{subsec:metVT}.

Secondly, we apply the method described above to real data from \textit{Kepler}. The result shows that the stellar variability plays a very important role in the detectability of the phase curves. Therefore, the best detection of phase curves occurs in the effective temperature range between 5500 and 6000 K (Sun like stars). In this range of effective temperature the variability is relatively stable compared to stars of spectral type $F$ or warmer, and $K$ or cooler where variability is increased by different mechanisms that result in a smaller fraction of recovered phase curves.

For the \textit{CHEOPS} telescope, where only specific events will be observed, it would take 9 complete orbits to detect a phase curve with 150 $ppm$ of amplitude on a star with magnitude $\sim$12 and a simple removal of variability. This is under the simple assumptions described in section \ref{sec:VT} and neglecting potential instrumental noises that can arise, and thus the number of needed orbits might be larger. One way to alleviate the problem of stellar variability in \textit{CHEOPS} data might be to intensively study the variability of the object from the ground while taking images from space with \textit{CHEOPS}. These observations should cover a longer period of time and be performed on a similar bandpass, in order to be able to fit a variability model to the star under study. This simple but costly ground-based observation time might improve the data taken from space, depending on the level of stellar variability, the amplitude of the phase curve, and the achievable precision from the ground.

The case of \textit{TESS} and \textit{PLATO} are different. These two missions will take photometric measurements in an uninterrupted manner. \textit{TESS} will monitor almost the whole sky, building light curves with a temporal scale of 27 days. Depending on the period, it will be possible to accumulate several orbits to better study phase curves. For \textit{PLATO}, its field will be the same for a long period of time (still under studying, but with a foreseen minimum of 2 years per long pointing direction) and that will allow much more orbits for a deeper study of exoplanetary phase curves. 

\begin{acknowledgements}
D.H. acknowledges the Spanish Ministry of Economy and Competitiveness (MINECO) for the financial support under the FPI program BES-2015-075200, R.A. under the Ramón y Cajal program RYC-2010-06519, and the three authors under the program RETOS ESP2014-57495-C2-R. We are grateful for the hospitality of J. Cabrera and the group of H. Rauer at the DLR Institute of Planetology, where part of this work has been performed.
\end{acknowledgements}

\medskip


\bibliographystyle{aa}
\bibliography{biblio}

\clearpage

\begin{appendix}
\section{Additional plots}\label{app:0}

\begin{figure}[h!]
	\resizebox{\hsize}{!}{\includegraphics{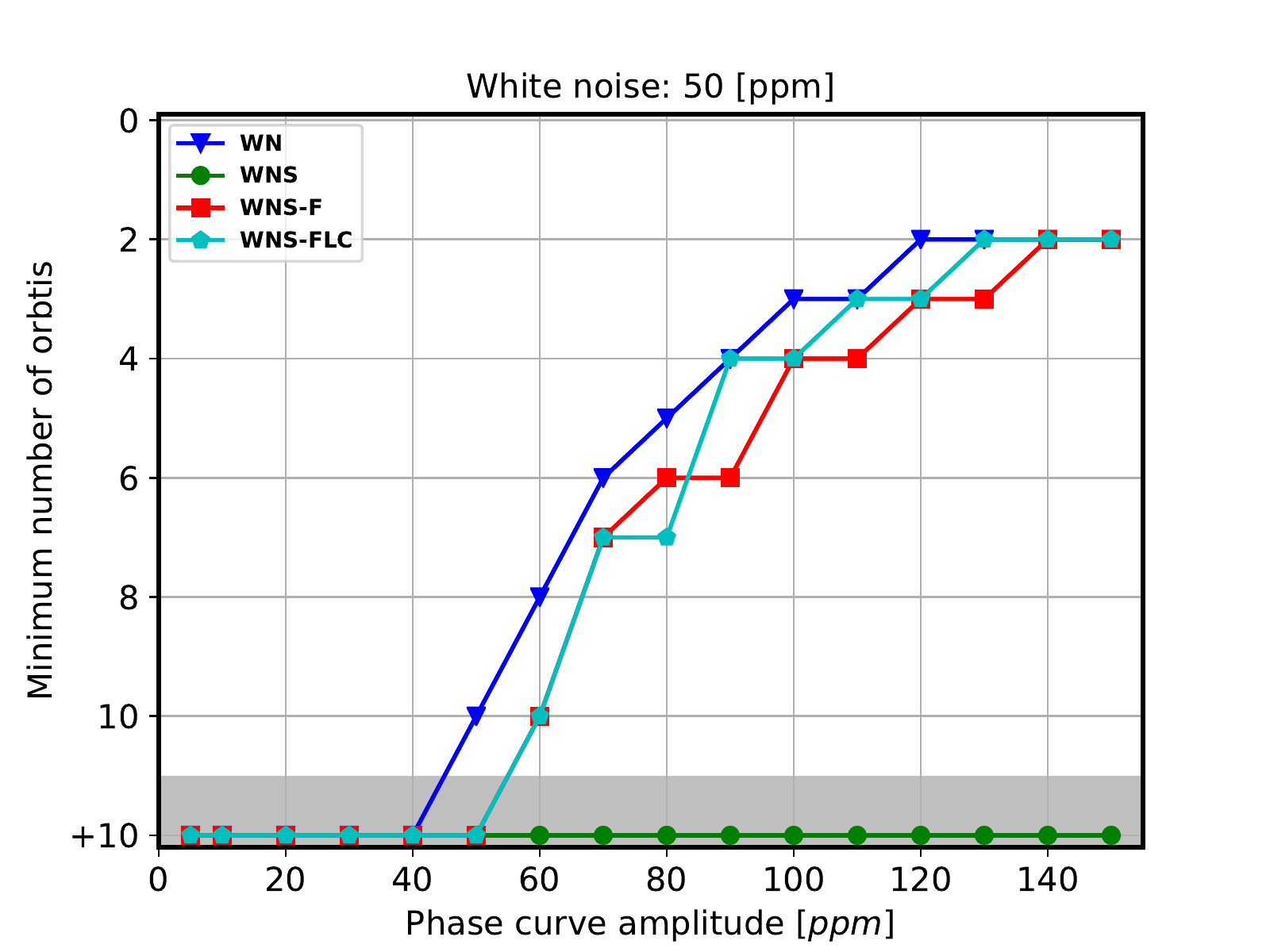}}
	\resizebox{\hsize}{!}{\includegraphics{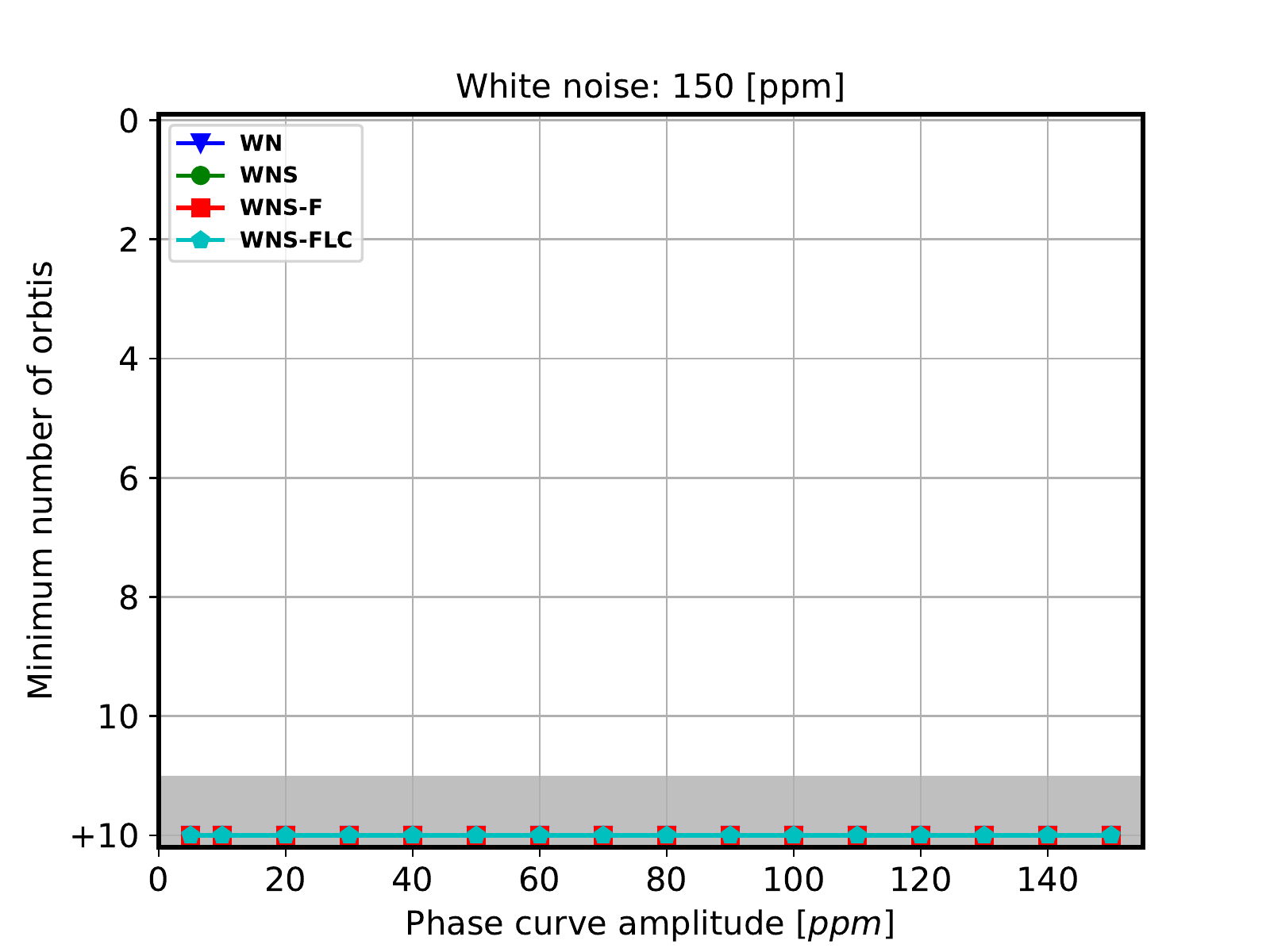}}
	\resizebox{\hsize}{!}{\includegraphics{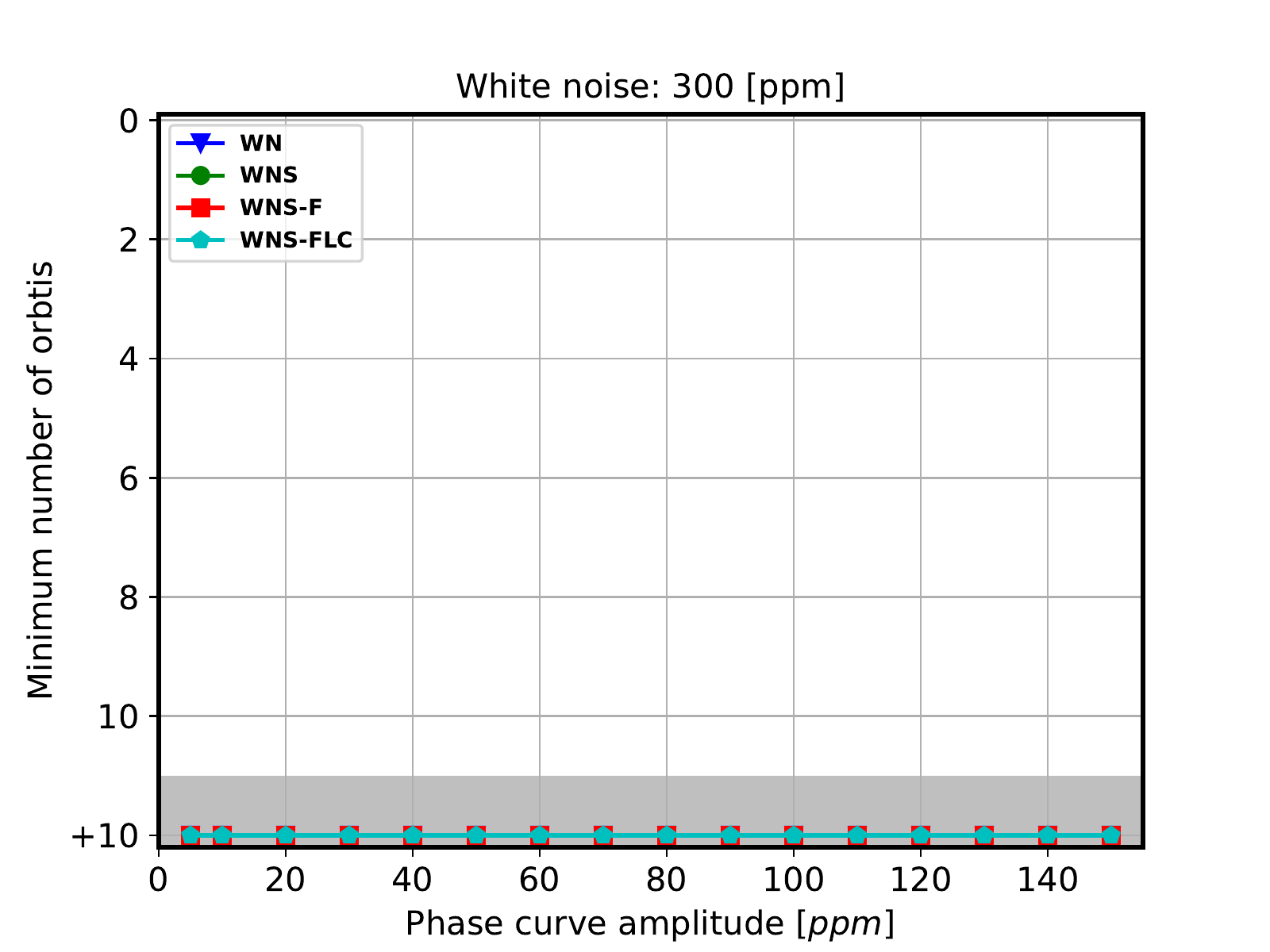}}
    \caption{Minimum number of orbits required to have significant phase curve detection, for different white noise (50, 150 and 300 ppm, top middle and bottom, respectively) and a period of 1.0 d.}
  	\label{fig:Re_1}
\end{figure}

\begin{figure}[h!]
	\resizebox{\hsize}{!}{\includegraphics{50ppm_2.pdf}}
	\resizebox{\hsize}{!}{\includegraphics{150ppm_2.pdf}}
	\resizebox{\hsize}{!}{\includegraphics{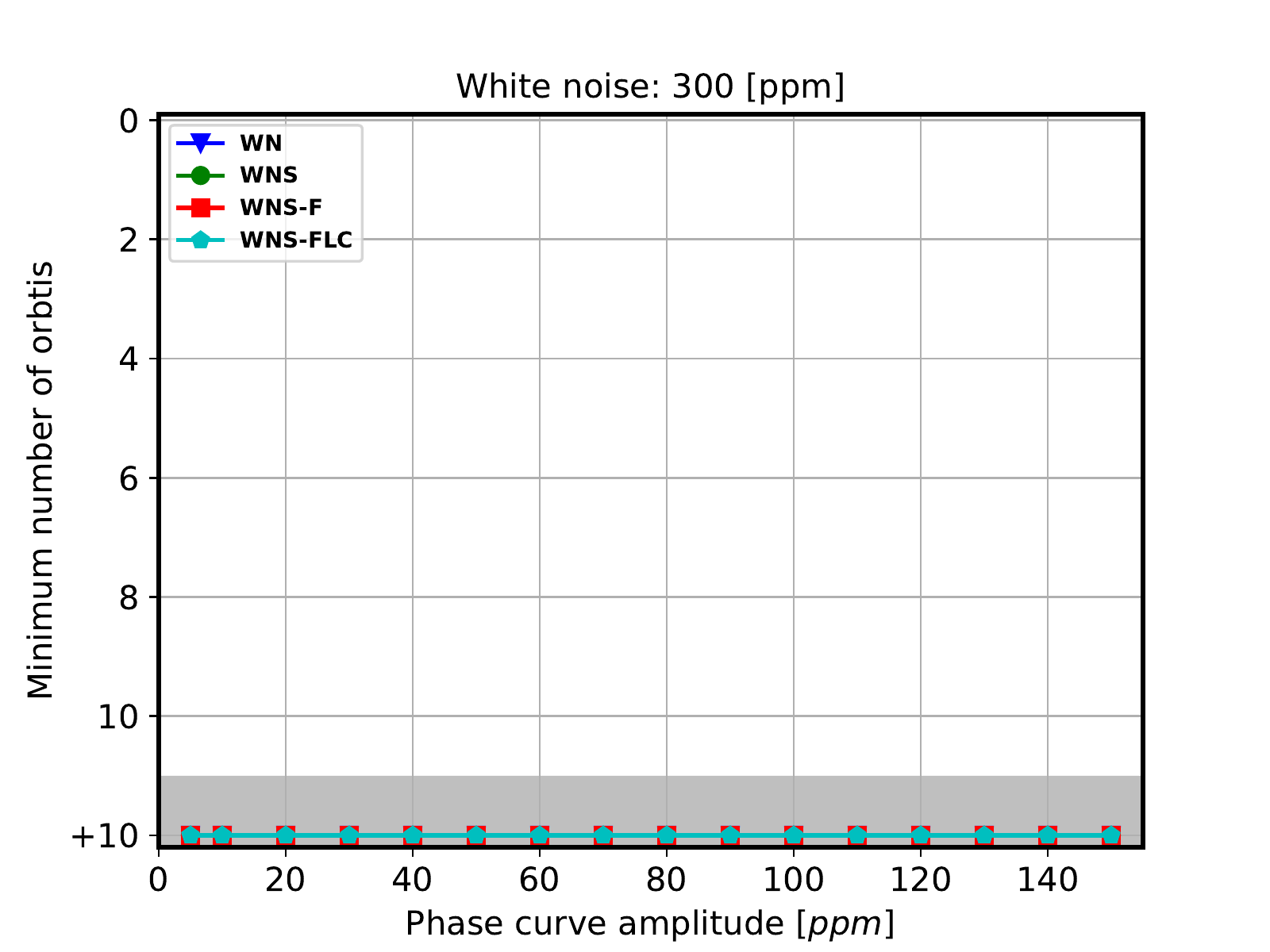}}
    \caption{Minimum number of orbits required to have significant phase curve detection, for different white noise (50, 150 and 300 ppm, top middle and bottom, respectively) and a period of 2.5 d.}
  	\label{fig:Re_2}
\end{figure}

\begin{figure}[h!]
	\resizebox{\hsize}{!}{\includegraphics{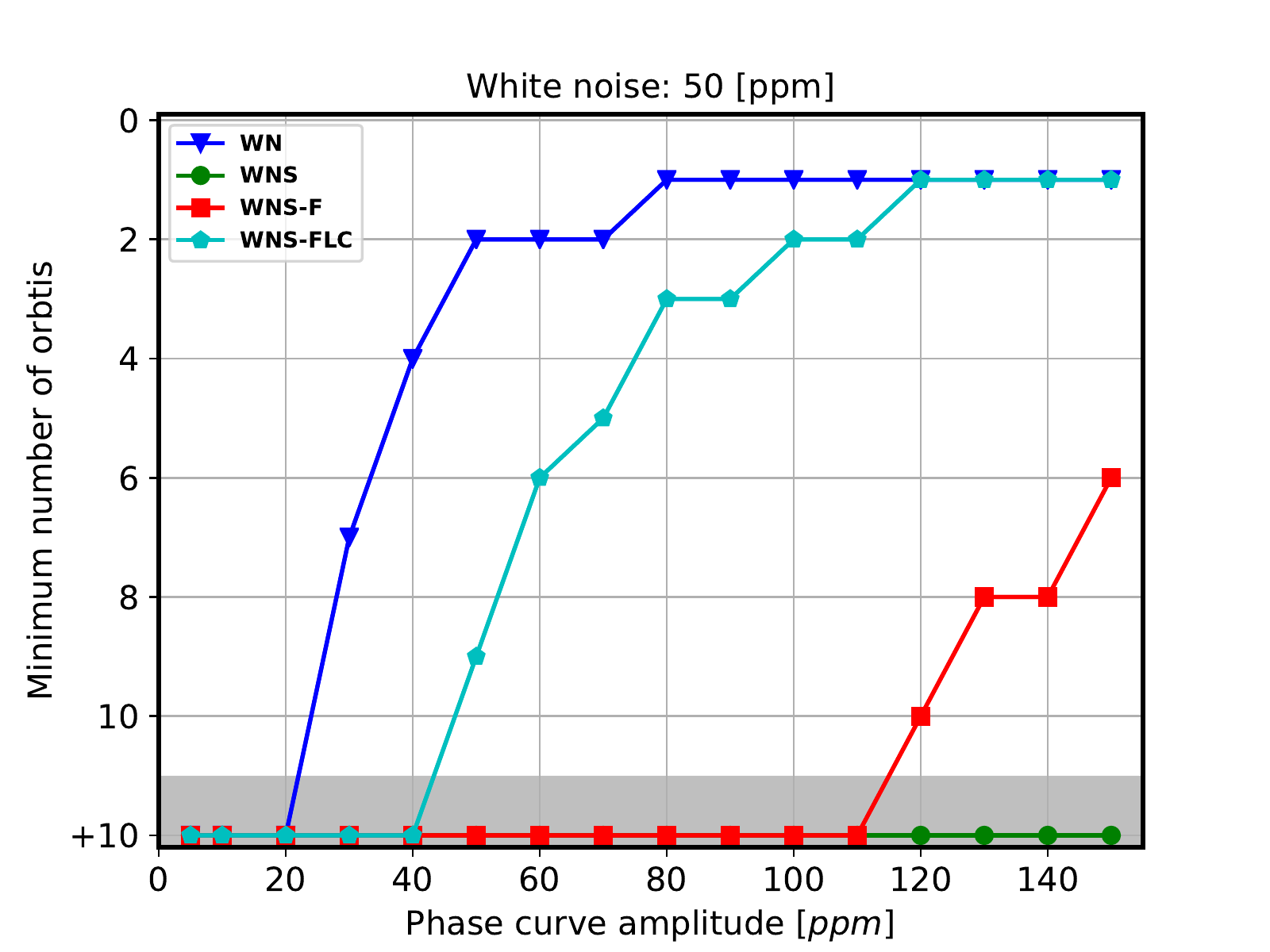}}
	\resizebox{\hsize}{!}{\includegraphics{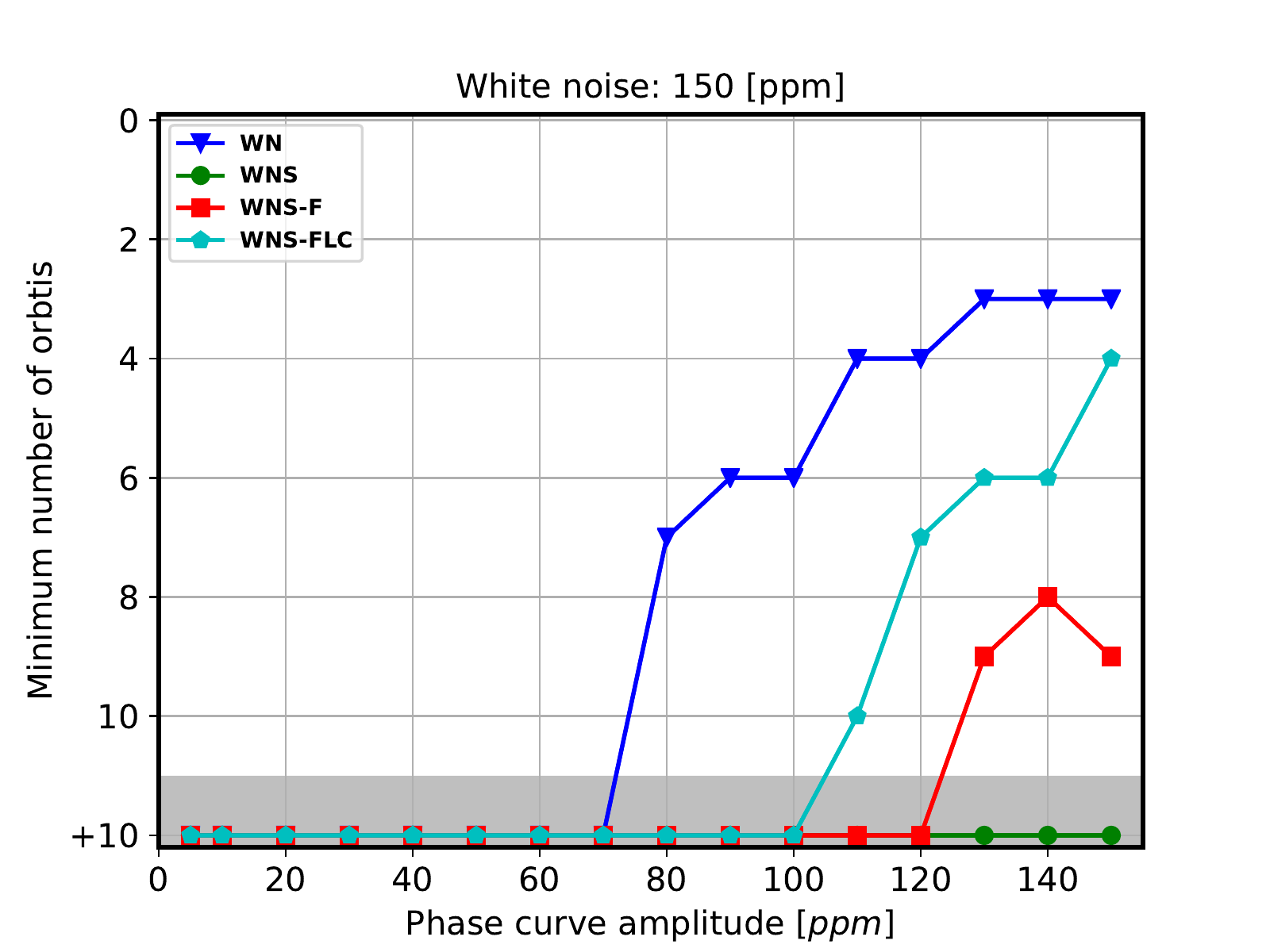}}
	\resizebox{\hsize}{!}{\includegraphics{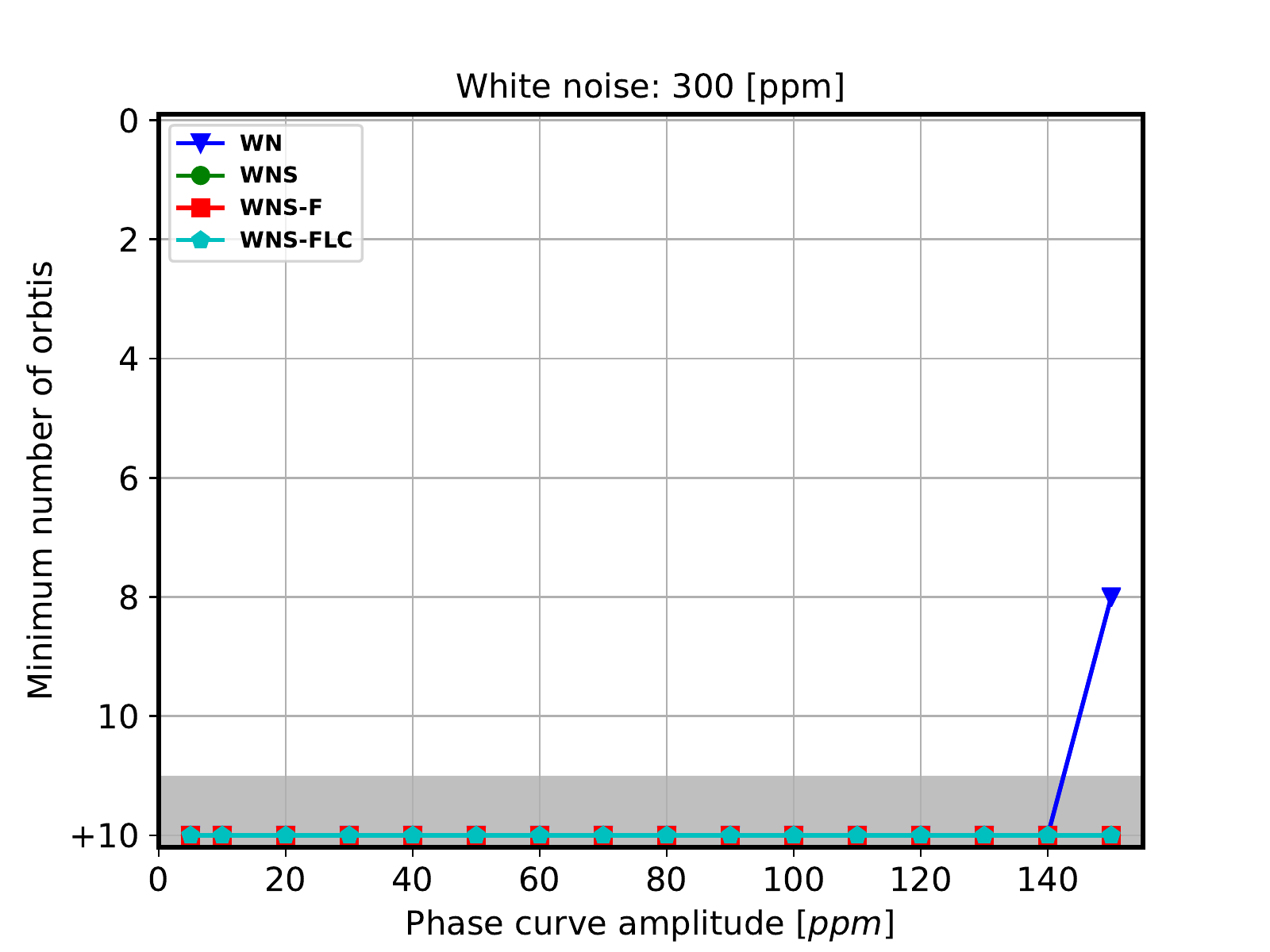}}
    \caption{Minimum number of orbits required to have significant phase curve detection, for different white noise (50, 150 and 300 ppm, top middle and bottom, respectively) and a period of 5.0 d.}
  	\label{fig:Re_5}
\end{figure}

\begin{figure}[h!]
	\resizebox{\hsize}{!}{\includegraphics{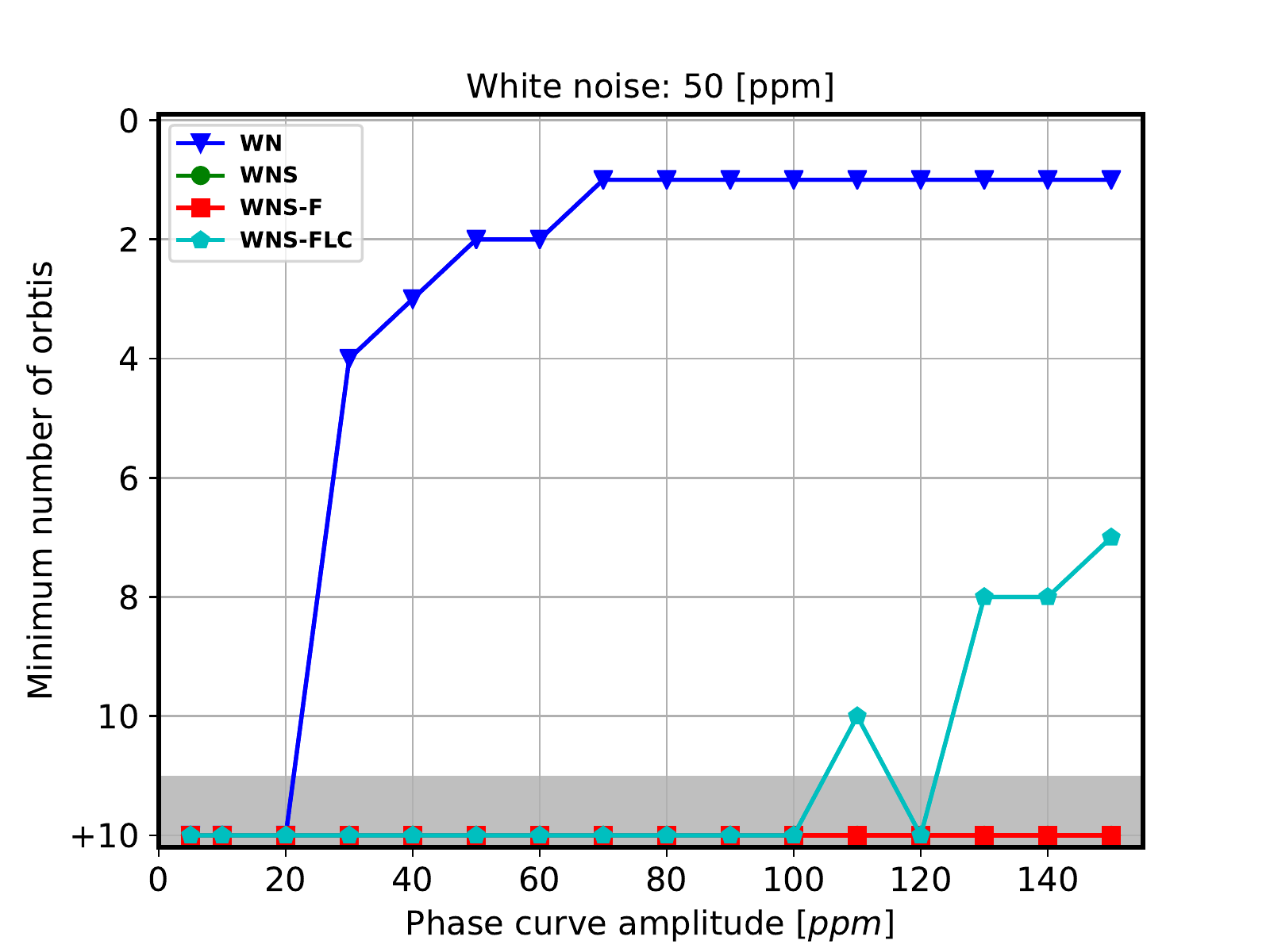}}
	\resizebox{\hsize}{!}{\includegraphics{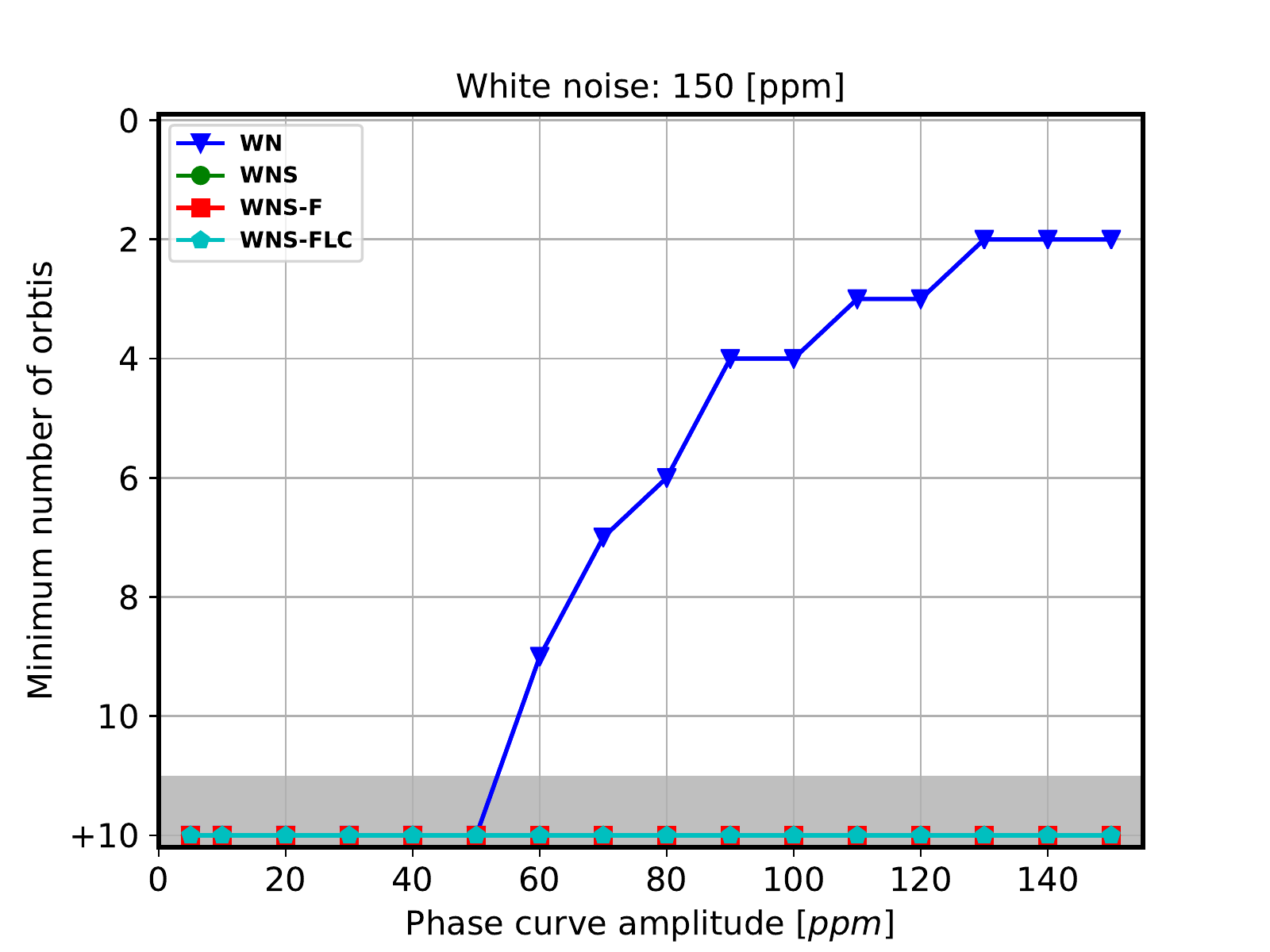}}
	\resizebox{\hsize}{!}{\includegraphics{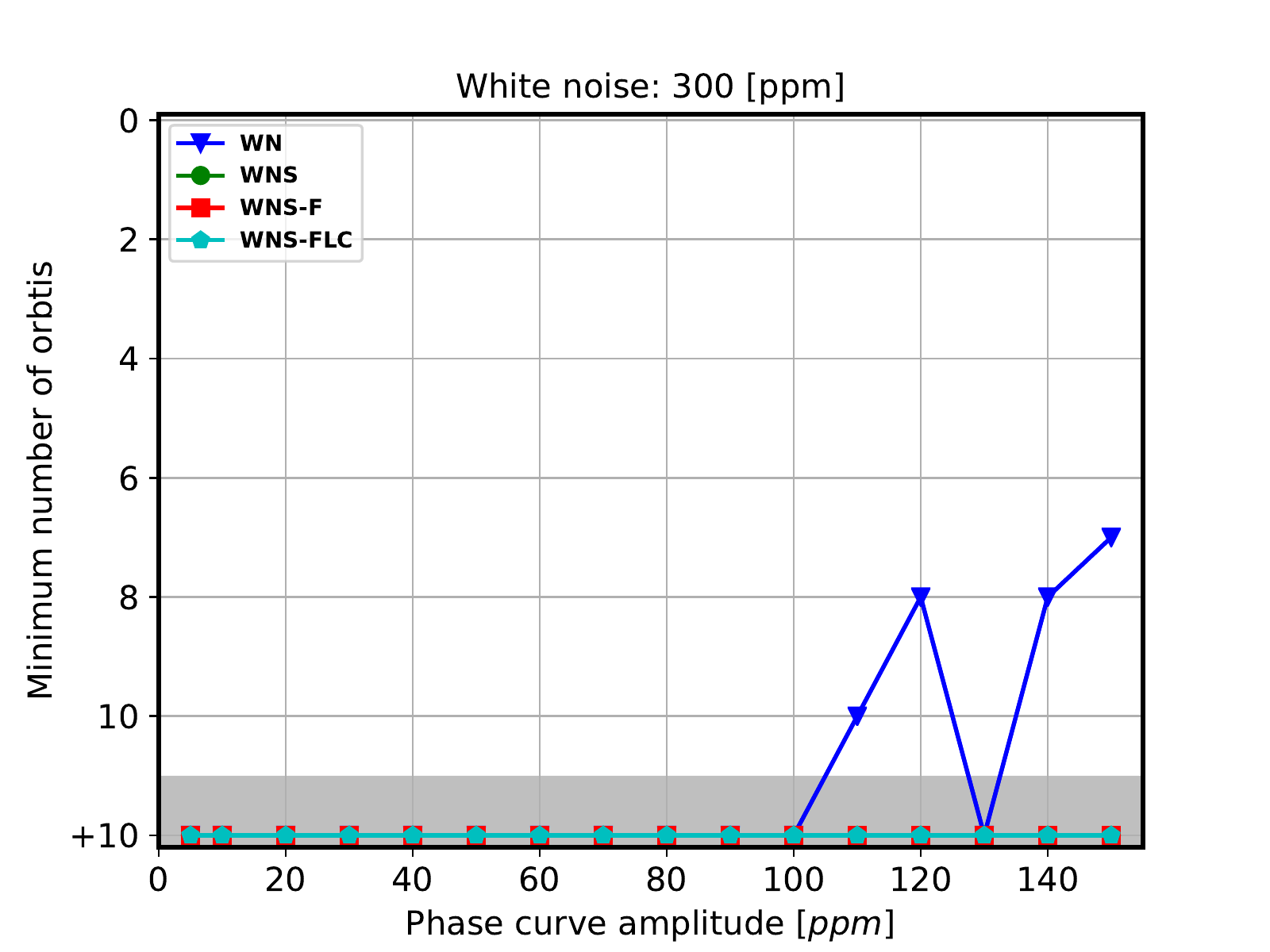}}
    \caption{Minimum number of orbits required to have significant phase curve detection, for different white noise (50, 150 and 300 ppm, top middle and bottom, respectively) and a period of 7.5 d.}
  	\label{fig:Re_7}
\end{figure}

\begin{figure}[h!]
	\resizebox{\hsize}{!}{\includegraphics{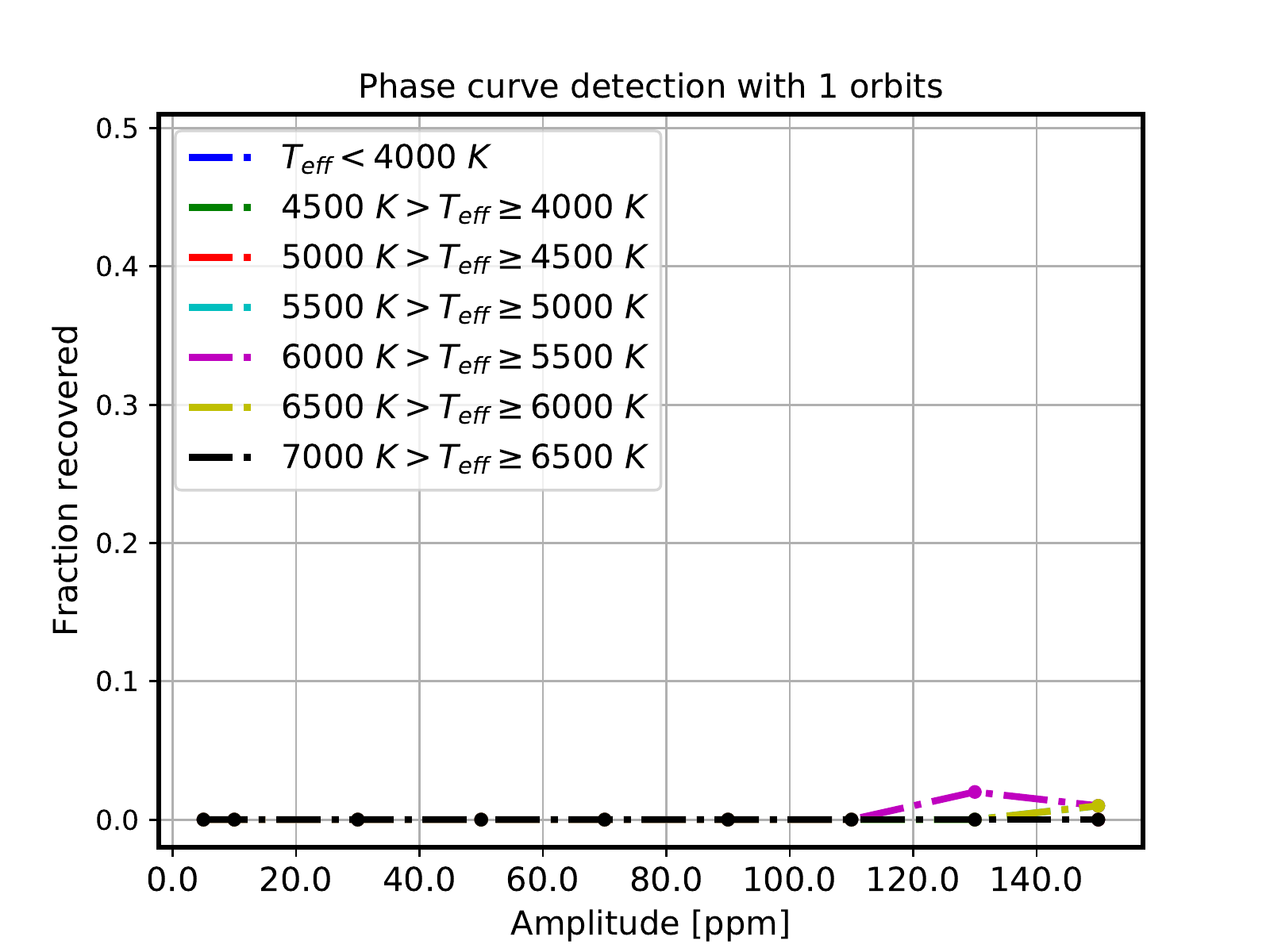}}
	\resizebox{\hsize}{!}{\includegraphics{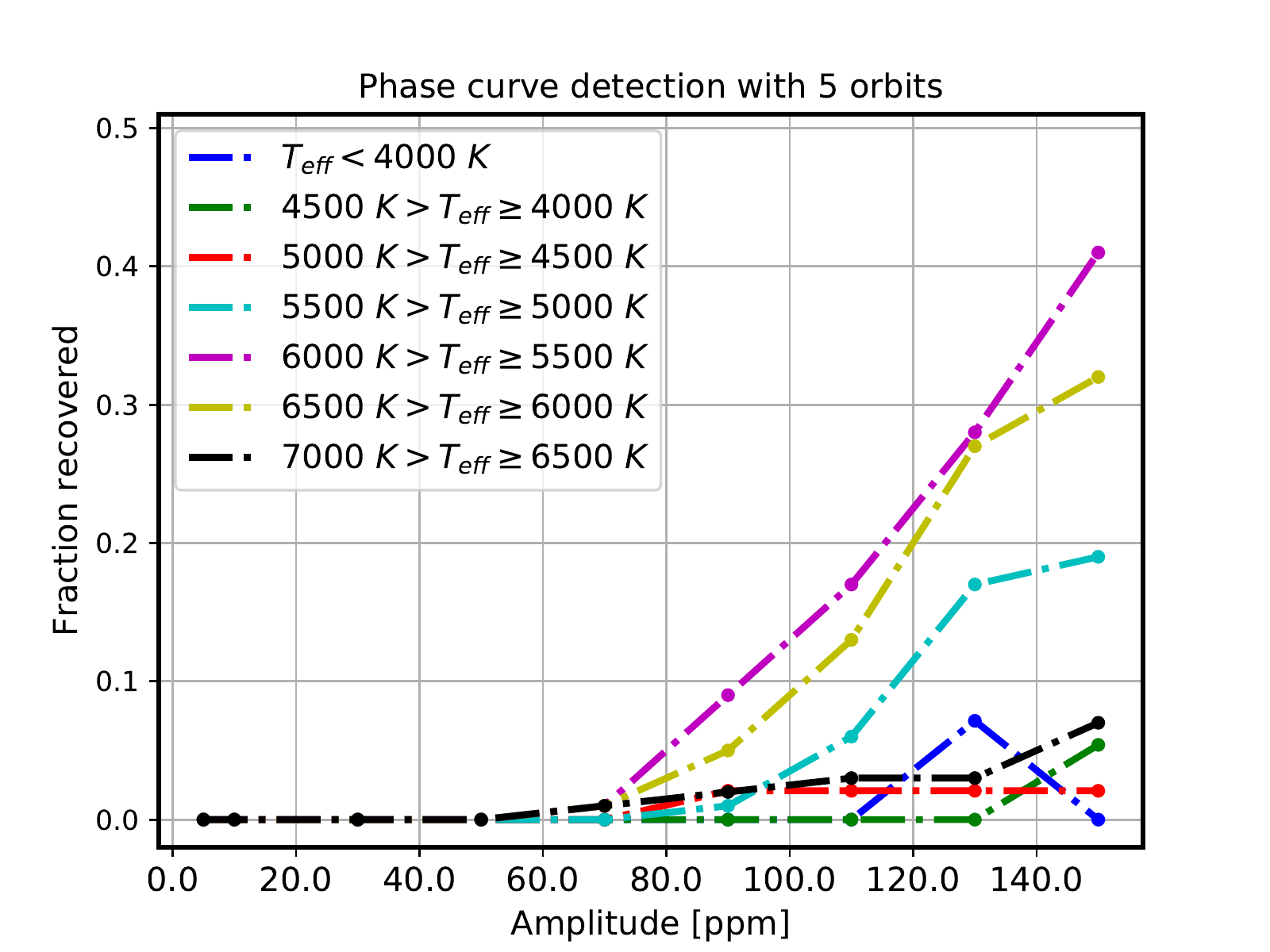}}
	\resizebox{\hsize}{!}{\includegraphics{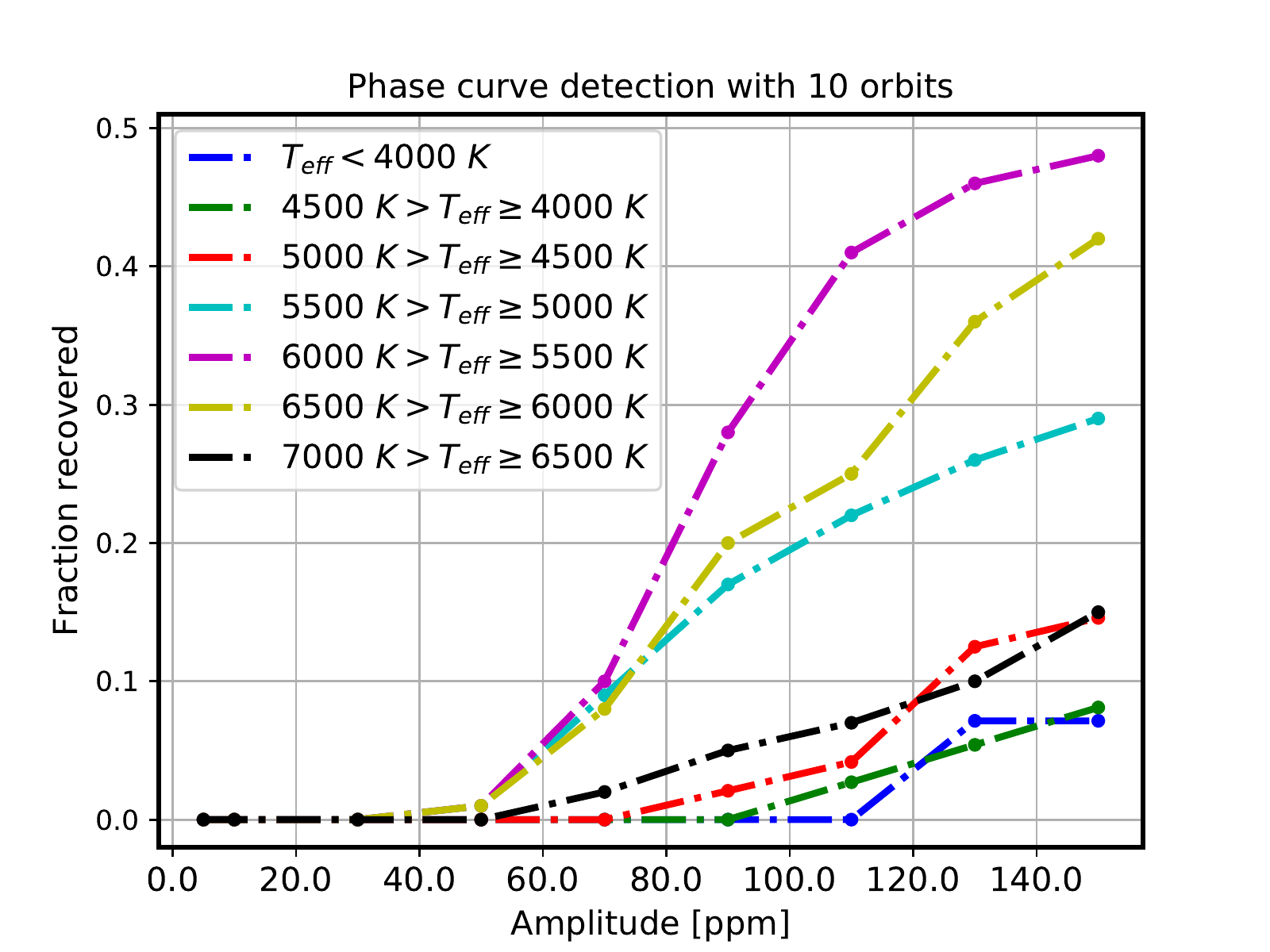}}
    \caption{Fraction of phase curves recovered as a function of the amplitude of the phase curve entered for a period of 1.0 d due to 1, 5 and 10 orbits considered (plots top, middle and bottom, respectively).}
  	\label{fig:Kep_N_1}
\end{figure}

\begin{figure}[h!]
	\resizebox{\hsize}{!}{\includegraphics{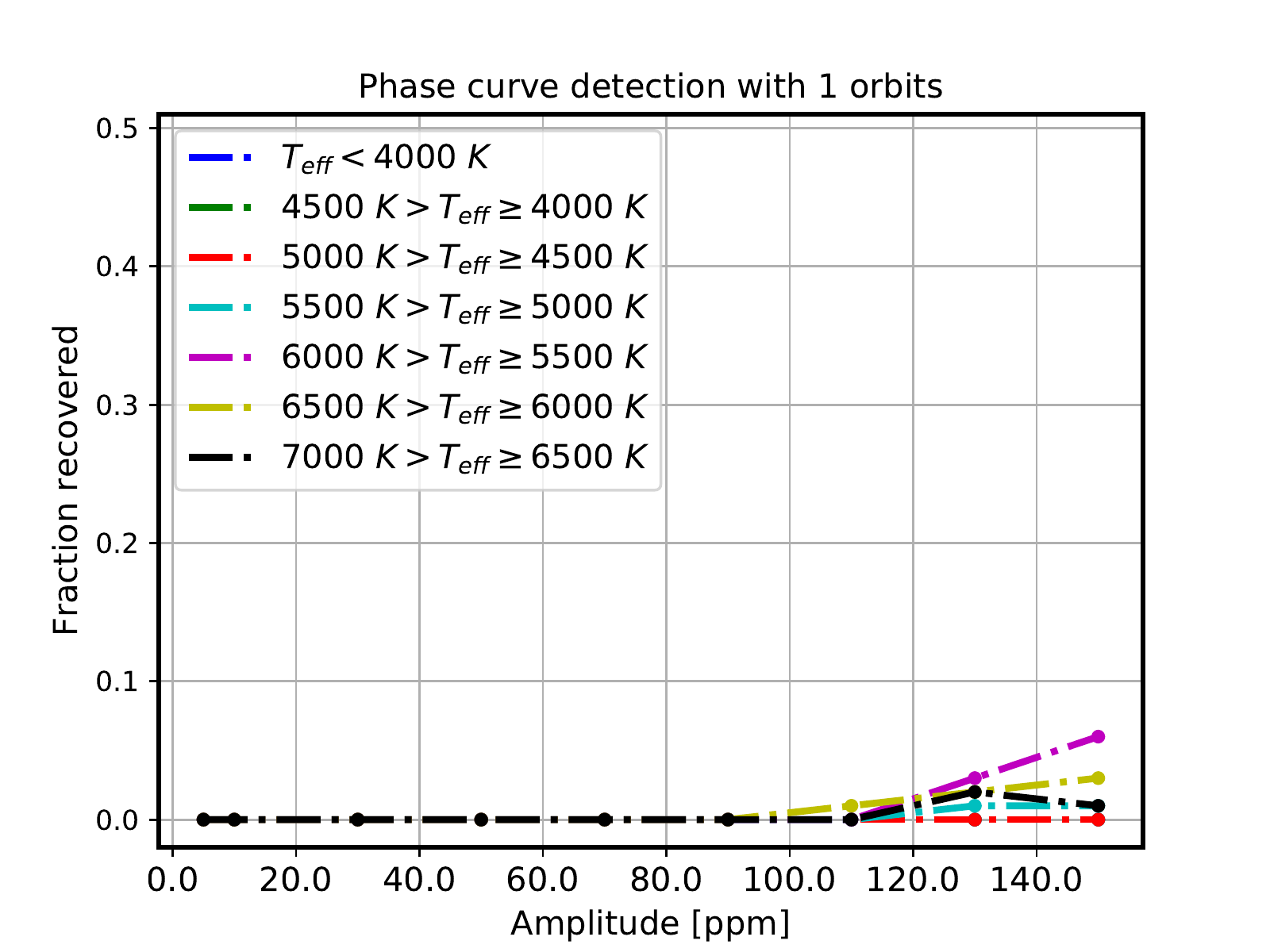}}
	\resizebox{\hsize}{!}{\includegraphics{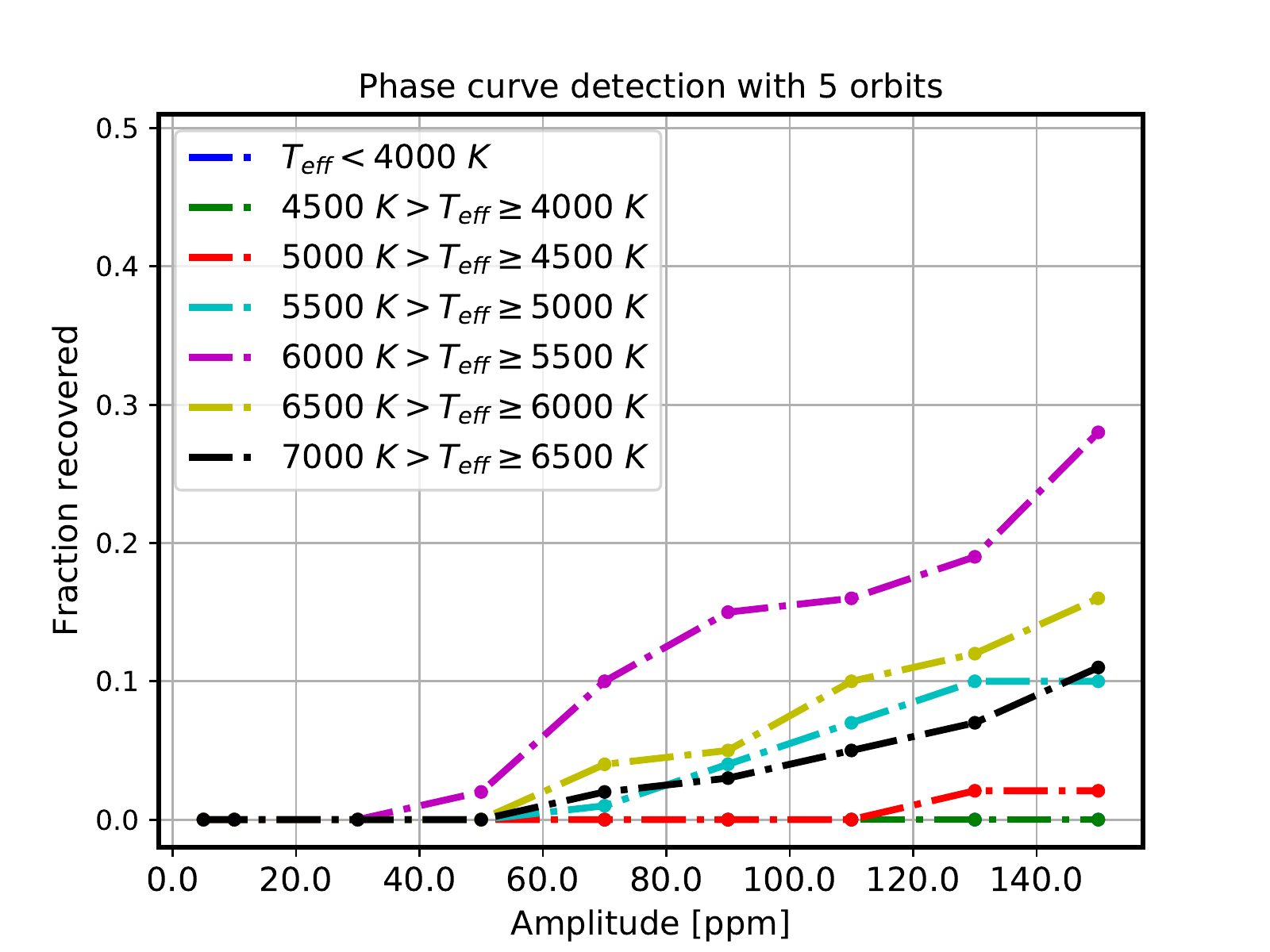}}
	\resizebox{\hsize}{!}{\includegraphics{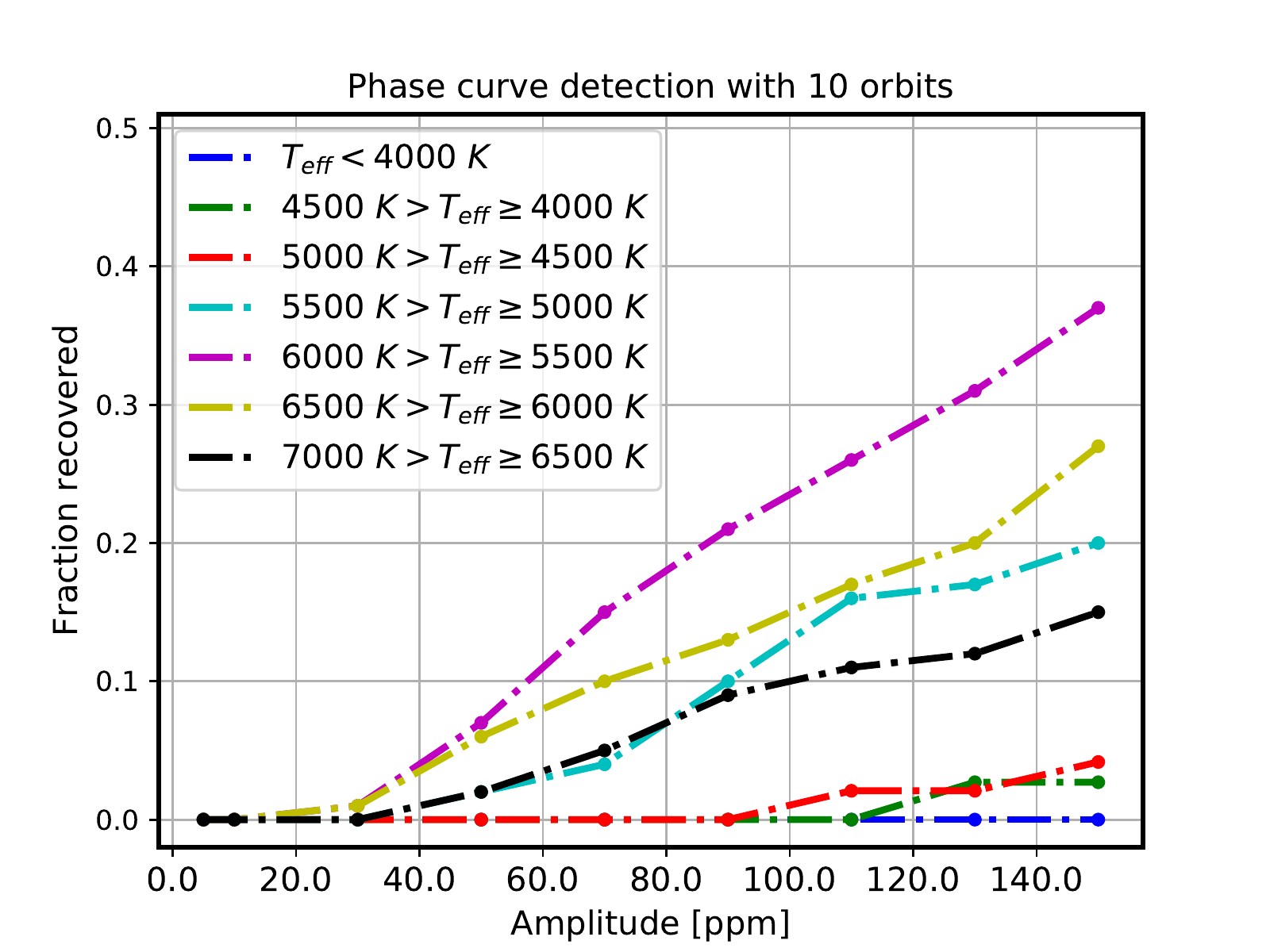}}
    \caption{Fraction of phase curves recovered as a function of the amplitude of the phase curve entered for a period of 5.0 d due to 1, 5 and 10 orbits considered (plots top, middle and bottom, respectively).}
  	\label{fig:Kep_N_5}
\end{figure}

\begin{figure}[h!]
	\resizebox{\hsize}{!}{\includegraphics{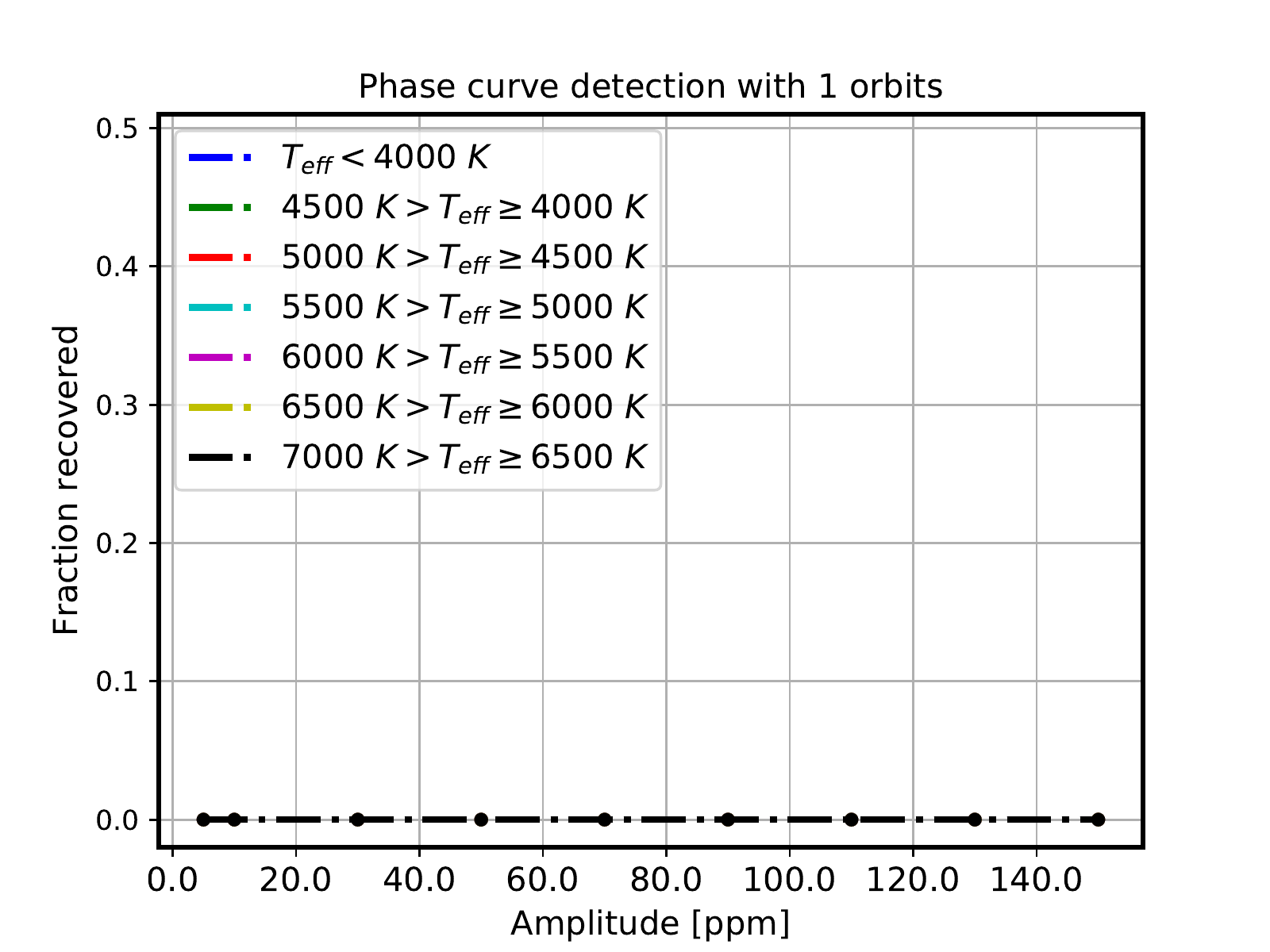}}
	\resizebox{\hsize}{!}{\includegraphics{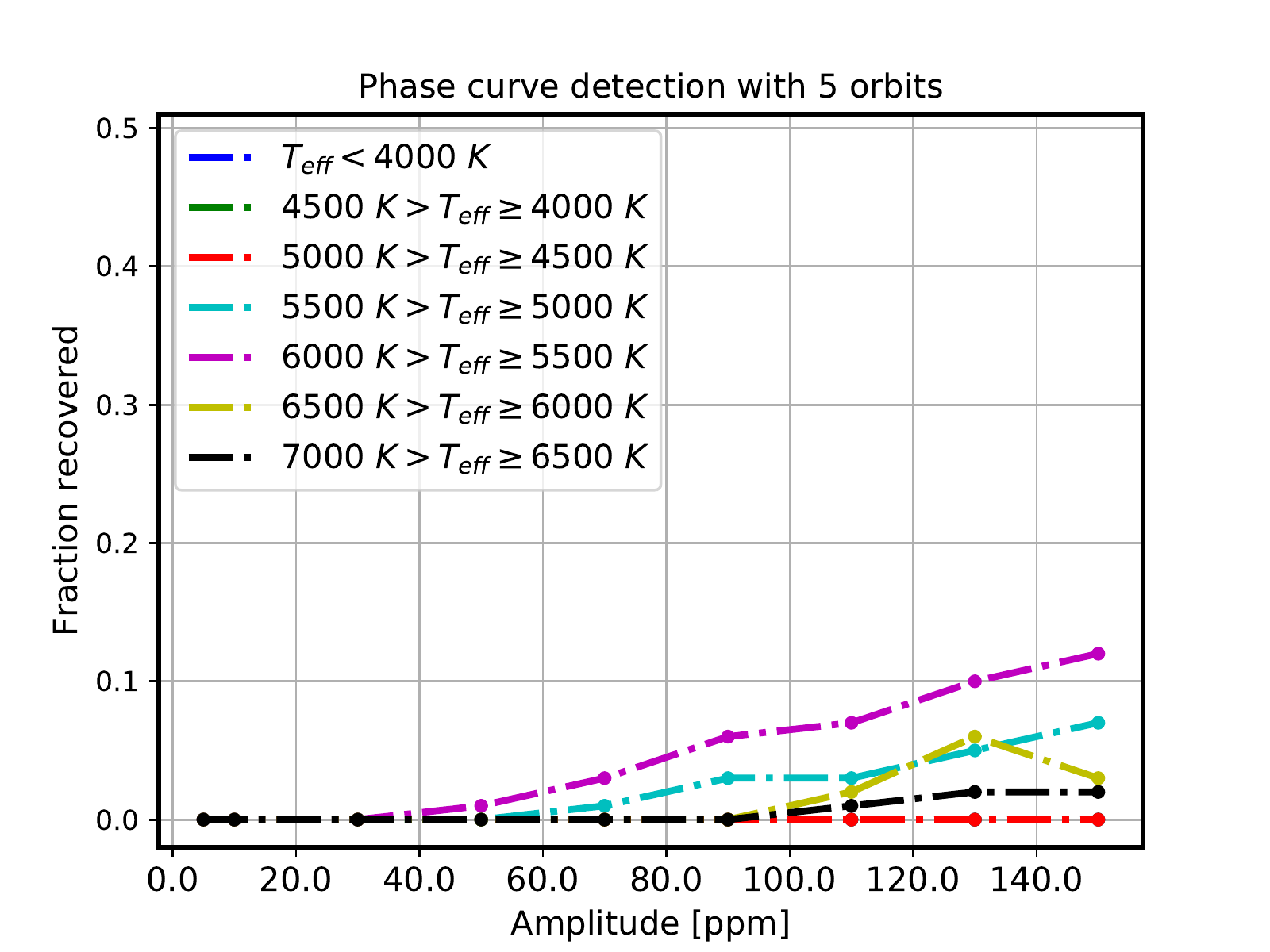}}
	\resizebox{\hsize}{!}{\includegraphics{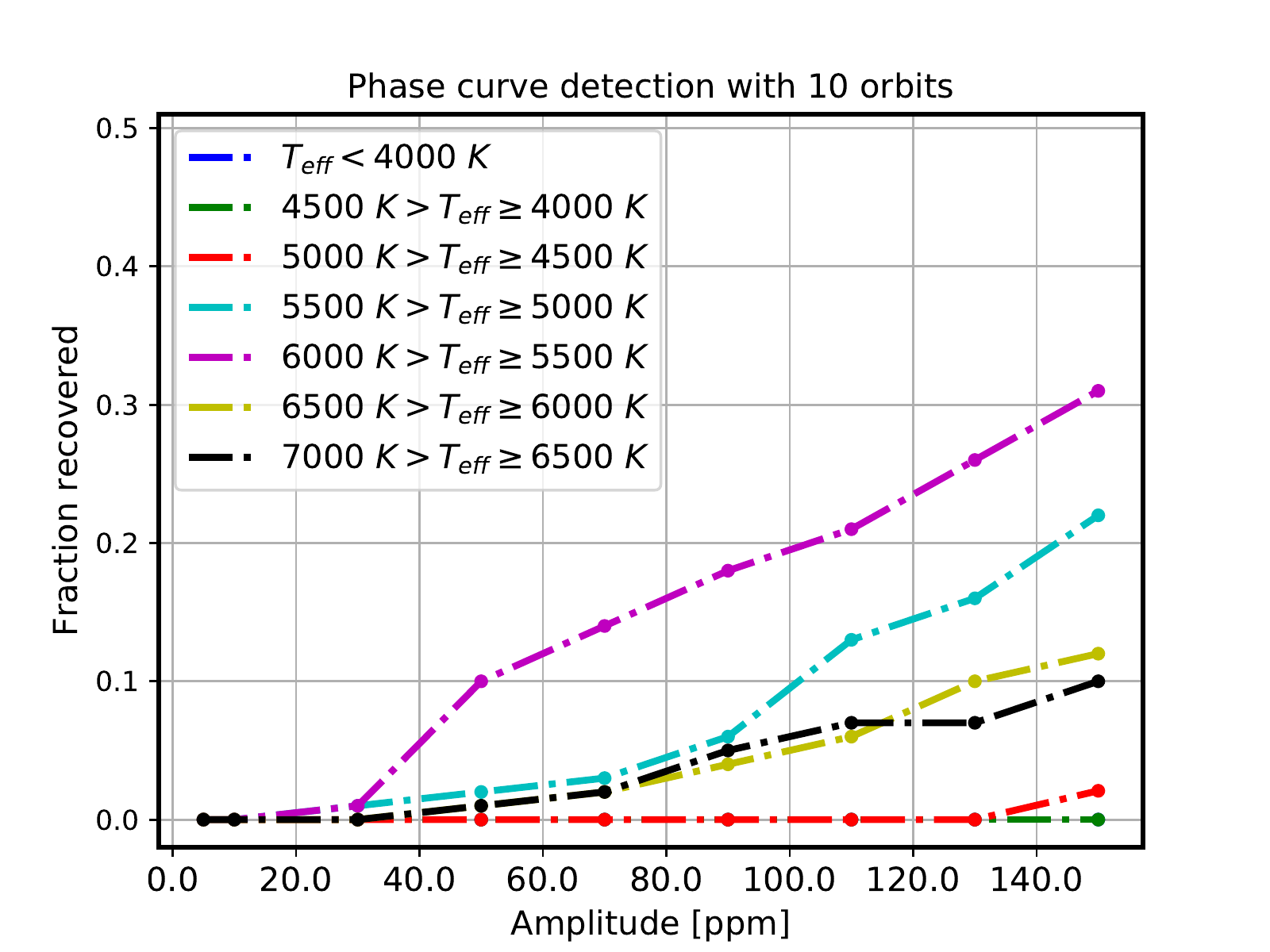}}
    \caption{Fraction of phase curves recovered as a function of the amplitude of the phase curve entered for a period of 7.5 d due to 1, 5 and 10 orbits considered (plots top, middle and bottom, respectively).}
  	\label{fig:Kep_N_7}
\end{figure}
\end{appendix}

\end{document}